\documentclass[prb,twocolumn,superscriptaddress,aps]{revtex4-2}
\usepackage[dvipsnames]{xcolor} % for text colors
\usepackage{amsfonts}
\usepackage{amsmath,amssymb,epsfig,color}
\usepackage{float}
\usepackage{amsmath}
\usepackage{amssymb}
\usepackage{url}
\usepackage{graphicx}%
\usepackage{dcolumn}%
\usepackage{bbold}
\usepackage{physics}
\usepackage{hyperref}
\begin{document}
\title{Planar Hall effect from superconducting fluctuations}
\author{L. Attias}
\affiliation{The Racah Institute of Physics, The Hebrew University of Jerusalem, Jerusalem 91904, Israel}
\author{K. Michaeli}
\affiliation{Department of Condensed Matter Physics, The Weizmann Institute of Science, Rehovot 76100, Israel}
\author{M. Khodas}
\affiliation{The Racah Institute of Physics, The Hebrew University of Jerusalem, Jerusalem 91904, Israel}
\begin{abstract}
We investigate the planar Hall effect (PHE) in two-dimensional (2D) superconductors with spin-orbit interactions, where transport anisotropy is induced by an in-plane magnetic field. While PHE typically arises from the breaking of basal mirror symmetry, when the field exclusively couples to spin degrees of freedom, it remains negligible in non-interacting systems. In this study, we explore anisotropic paraconductivity as an alternative mechanism for PHE observed in 2D superconductors in the normal state. Due to the momentum dependence of spin-orbit interactions, the field-induced pair breaking exhibits anisotropy. To elucidate this phenomenon, we compute the PHE for the Rashba spin-orbit interaction. Our analysis reveals that Cooper pairs propagating along the field experience stronger pair breaking compared to those moving perpendicular to the field. This physical insight is corroborated by explicit calculations of paraconductivity.
\end{abstract}
\maketitle
\section{Introduction}
\label{sec:Intro}

The Planar Hall effect (PHE) is a distinct kind of a magneto-resistance anisotropy induced by 
magnetization and/or the applied magnetic field $\mathbf{B}$.
The PHE implies the disparity in conductivity between the current flowing parallel (\( \sigma_{\parallel}\)) and perpendicular (\(\sigma_{\perp}\)) to the magnetization or in-plane magnetic field ($\mathbf{B}$) \cite{Goldberg1954,Zhong2023}.
This effect has been observed in LaAlO$_3$/SrTiO$_3$ and 
LaVO$_3$/KTaO$_3$ interfaces \cite{Annadi2013,Rout2017,Maniv2017,Wadehra2020}, in topological insulator nano-devices \cite{Sulaev2015,Taskin2017,Bauer2017,Mehraeen2023}, semiconducting thin films \cite{Akouala2019}, superconductors with strong spin-orbit coupling \cite{Li2022}, kagome metals \cite{Li2023}, and Weyl semimetals \cite{Onofre2023}.

The standard Hall effect arises from the Lorentz force acting on charge carriers. 
This effect is contained in the antisymmetric part of the conductivity tensor, denoted as $\hat{\sigma}$. According to the Onsager relation, $\hat{\sigma}_{xy}(\mathbf{B}) = \hat{\sigma}_{yx}(-\mathbf{B})$, indicating that the standard Hall current is odd in $\mathbf{B}$.

The PHE, in contrast, is contained in the symmetric part of $\hat{\sigma}$, which is even in $\mathbf{B}$. 
As a result, in magnetic materials, PHE is sensitive to the square of magnetization rather than to the magnetization itself. 
This characteristic makes PHE a valuable tool for detecting anti-ferromagnetic transitions \cite{Bodnar2018,Baltz2018,Yin2019}.

\begin{figure}
    \centering
    \includegraphics[scale = 0.59]{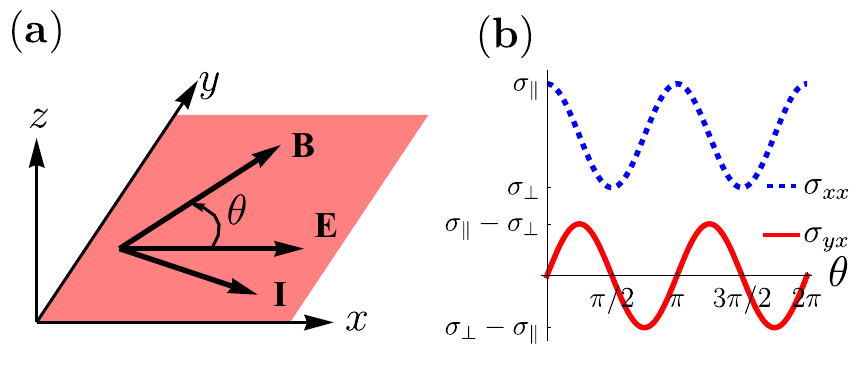}
    \caption{
	    (a) 
        The 2D system (shaded area) in the planar configuration.
        The system, the electric field (\( \vb{E} \parallel \hat{x} \)), magnetic field  (\( \vb{B} \)) and the current (\( \vb{I} \)) are in the \(xy\)-plane. 
        %The 2D system is indicated by the shaded (pink) rectangle. 
        \( \vb{E} \) and  \( \vb{B} \) form the angle \( \theta \). 
        (b) %Angular behavior of the PHE. 
        The dashed (blue) line is the longitudinal conductance %along the direction of the electric field 
        \( \sigma_{xx} \left( \theta \right)\), and the solid (red) line is the Hall conductance \( \sigma_{yx} \left( \theta \right) \).  
        The PHE implies $\pi$-periodic angular dependence of the conductance tensor with the Hall conductance, \( \sigma_{yx} \left( \theta \right) = \sigma_{xy} \left( \theta \right) =  (\sigma_{\parallel} - \sigma_{\perp}) \sin( 2 \theta ) \), resulting from a finite anisotropy, $\sigma_{\parallel} \neq \sigma_\perp$.
    }
    \label{fig:plane}
\end{figure}

In this study, we concentrate on two-dimensional (2D) systems confined to the basal ($xy$) plane. 
We consider the planar configuration where both electric and magnetic fields lie in-plane, as illustrated in Fig.~\ref{fig:plane}(a).
In this geometry the Lorentz force acts out-of-plane, and therefore causes no current.
Consequently, the conventional Hall conductivity is zero. 
As a result, the 2D conductivity tensor is symmetric and even in $\mathbf{B}$.
Furthermore, the conductivities $\sigma_{\parallel,\perp}$ are principle values of this tensor.

The preceding arguments elucidate why the PHE is frequently investigated in the planar configuration \cite{Li2024}. 
In one realization of this setup, the 2D system is formed at the (111) interface between LaAlO$_3$ and SrTiO$_3$ \cite{Annadi2013,Rout2017,Maniv2017}.
These systems are inherently anistropic owing to the underlying  crystal structure.
To distinguish this anisotropy from the PHE, the conductivity is monitored as the field $\mathbf{B}$ rotates in the (111) plane.
The hexagonal symmetry of the interface results into a six-fold angular variation of $\hat{\sigma}$.
In contrast, the field-induced anisotropy in the form of the PHE manifests as a distinct two-fold angular variation of \( \hat{\sigma} \).
As the field rotates through an angle \( \theta \) relative to a fixed coordinate frame, the PHE implies a finite Hall conductivity, \( \sigma_{xy}(\theta) = (\sigma_{\parallel} - \sigma_{\perp}) \sin 2 \theta \), satisfying \( \sigma_{xy}(\theta) =\sigma_{yx}(\theta) \). Simultaneously, the diagonal elements acquire angular variations, \( \delta  \sigma_{xx,yy} = \pm (\sigma_{\parallel} - \sigma_{\perp}) \cos 2 \theta /2 \), as illustrated in Fig.~\ref{fig:plane}b.

%new text
The PHE is distinct from the anisotropy induced by external symmetry breaking perturbations \cite{Haim2022} such as e.g. strain \cite{Hamill2021} or anisotropic magnetic impurities \cite{Wickramaratne2021}.
In the case of the PHE, the anisotropy is induced by the applied magnetic field and is an intrinsic property of the system.

Another representative experiment reports the two-fold variation of the thermodynamic and transport properties of a few-layer NbSe$_2$ as the in-plane field rotates \cite{Hamill2021}. In a monolayer of the same material on a substrate, the two-fold variation is superimposed on the six-fold variation expected for hexagonal crystals, at least within certain range of applied fields \cite{Cho2022}.

Theoretically the PHE has been investigated in 2D spin-orbit coupled systems \cite{Schwab2002,Raimondi2005}.
The conclusion of these studies is that PHE vanishes unless the Zeeman splitting induced by the in-plane magnetic field exceeds the spin splitting caused by the spin-orbit interaction (\(\Delta_{\mathrm{SO}} \)). 
The spin-orbit interaction introduces the anomalous velocity, a spin-dependent term, to the current operator.
The anomalous velocity and the vertex corrections to the normal velocity cancel each other out \cite{Sinova2004,Inoue2004,Rashba2004,Raimondi2005,Sinova2015}.
This explains the vanishing of the PHE in non-interacting spin-orbit coupled 2D systems.

The theoretical findings mentioned above appear to conflict with the observation of a finite PHE. To reconcile these disparities, in this study, we propose and examine an alternative, yet universal mechanism of the PHE. We observe that both the LaAlO$_3$/SrTiO$_3$ interfaces and the few-layer NbSe$_2$ on a substrate exhibit superconductivity upon cooling to a finite critical temperature (\(T_c\)). Furthermore, both families of systems break the mirror symmetry of the basal plane (\(\sigma_h\)).
In light of these observations, we make two key assumptions. 
%Firstly, we consider a finite temperature for the superconducting transition, denoted as \(T_c\). 
Firstly, we assume the system has a finite $T_c$.
Secondly, we focus on 2D systems with broken $\sigma_h$ and inversion symmetry (\(\mathcal{P}\)), ensuring the existence of finite spin splitting of electron bands due to the spin-orbit interaction.

We demonstrate that in a 2D superconductor with broken \( \sigma_h \) symmetry, the conductivity arising from superconducting fluctuations (paraconductivity) \cite{Larkin2005} becomes anisotropic in the presence of an in-plane magnetic field. 
We provide an explicit microscopic calculation of the PHE in such system.
Importantly, since the PHE vanishes in the non-interacting limit, the anisotropic paraconductivity alone explains the PHE for a large range of temperatures, not necessarily confined to those close to \( T_c \).

The paper is structured as follows.
Section \ref{sec:phenomenology} provides a phenomenological description of the PHE, grounded in symmetry constraints on the Cooper pair dispersion relation.
In Sec.~\ref{sec:micro_model}, we introduce the microscopic model used to investigate the PHE.
A summary of the results obtained within this model is presented in Sec.~\ref{sec:summary_disp}.
The detailed calculations leading to these results are provided in Sec.~\ref{sec:calculation}, where we compute the coefficients in the Ginzburg-Landau free energy up to second order in Cooper pair momentum.
Finally, Sec.~\ref{sec:conclusion} discusses the results and outlines directions for future research.

\section{PHE from paraconductivity}
\label{sec:phenomenology}

In this section we delve into the phenomenology of the PHE arising from superconducting fluctuations.
We find as one of our main results, 
\begin{align}\label{eq:res}
    \sigma_\parallel - \sigma_\perp = \sigma_{\mathrm{AL}} B^2 \frac{\mathcal{L}_2}{\xi_0^2} \, .
\end{align}
The PHE \eqref{eq:res} is proportional to the zero field Aslamazov-Larkin paraconductivity, $\sigma_{\mathrm{AL}}$. 
In two dimensions the latter is universal, $\sigma_{\mathrm{AL}} = e^2/ 16 \epsilon$, where
$\epsilon = (T - T_c)/T_c$, \cite{Aslamazov1968}.  
In what follows we use units with $\hbar = c = k_B =1$.
Hence, the PHE \eqref{eq:res} is determined by two phenomenological constants.
The first is the standard zero-field Cooper pair size, $\xi_0$, while the second parameter, $\mathcal{L}_2$, characterizes the spatial anisotropy of the Cooper pairs. 
We further elucidate the physical significance of both parameters and derive Eq.~\eqref{eq:res} using the phenomenological time-dependent Ginzburg-Landau approach \cite{Dorsey1991}.

The paraconductivity is mediated by the Cooper pairs carrying a \(2e\) charge.
In the normal state, these Cooper pairs have a finite lifetime $\gamma_{\text{GL}}^{-1}$.
They manifest as classical fluctuations of the order parameter $\Psi(\mathbf{x})$, where $\mathbf{x}$ represents a spatial coordinate. 
The fluctuations, characterized by momentum $\mathbf{q} = (q_x, q_y)$, are expressed as $\Psi(\mathbf{x}) = \Psi_{\mathbf{q}} e^{ i \mathbf{q} \mathbf{x} }$.
The Ginzburg-Landau free energy $F_{\mathrm{GL}}$ consists of terms up to second and fourth order in $\Psi(\mathbf{x})$.
Above $T_c$, we disregard the fourth-order terms, leading to $F_{\mathrm{GL}} = \sum_{\mathbf{q}} |\Psi_{\mathbf{q}}|^2 \varepsilon(\mathbf{q})$, where $\varepsilon(\mathbf{q})$ denotes the dispersion relation of the Cooper pairs.
This dispersion relation, $\varepsilon(\mathbf{q})$, in turn defines the Cooper pair velocity $\mathbf{v}_{\mathbf{q}} = \partial_{\mathbf{q}} \varepsilon(\mathbf{q})$.

Within the framework of time-dependent Ginzburg-Landau theory, the paraconductivity is given by \cite{Larkin2005}
\begin{align}\label{eq:conductivity}
	\delta \sigma_{\alpha \beta}
	= 2 e^2 T \gamma_{\text{GL}} \sum_{ \mathbf{q} } \frac{ v^\alpha_{\mathbf{q}} 
 v^\beta_{\mathbf{q}} }{ \left[\varepsilon(\mathbf{q})\right]^3} \, .
\end{align}
Clearly, the tensor in Eq.~\eqref{eq:conductivity} is isotropic if the pair dispersion is, $\varepsilon(\mathbf{q})= \varepsilon(q)$.
Therefore, in the proposed scenario the transport anisotropy arises from the anisotropy of the pair dispersion $\varepsilon(\mathbf{q})$.
Furthermore, similar to the standard case of Aslamazov-Larkin corrections, the dominant contribution to the conductivity originates from small momenta of the Cooper pairs on the order of the inverse coherence length. 
Consequently, it suffices to retain terms up to second order in $\mathbf{q}$ in $\varepsilon(\mathbf{q})$.

\begin{figure}
    \centering
    \includegraphics[scale = 0.3]{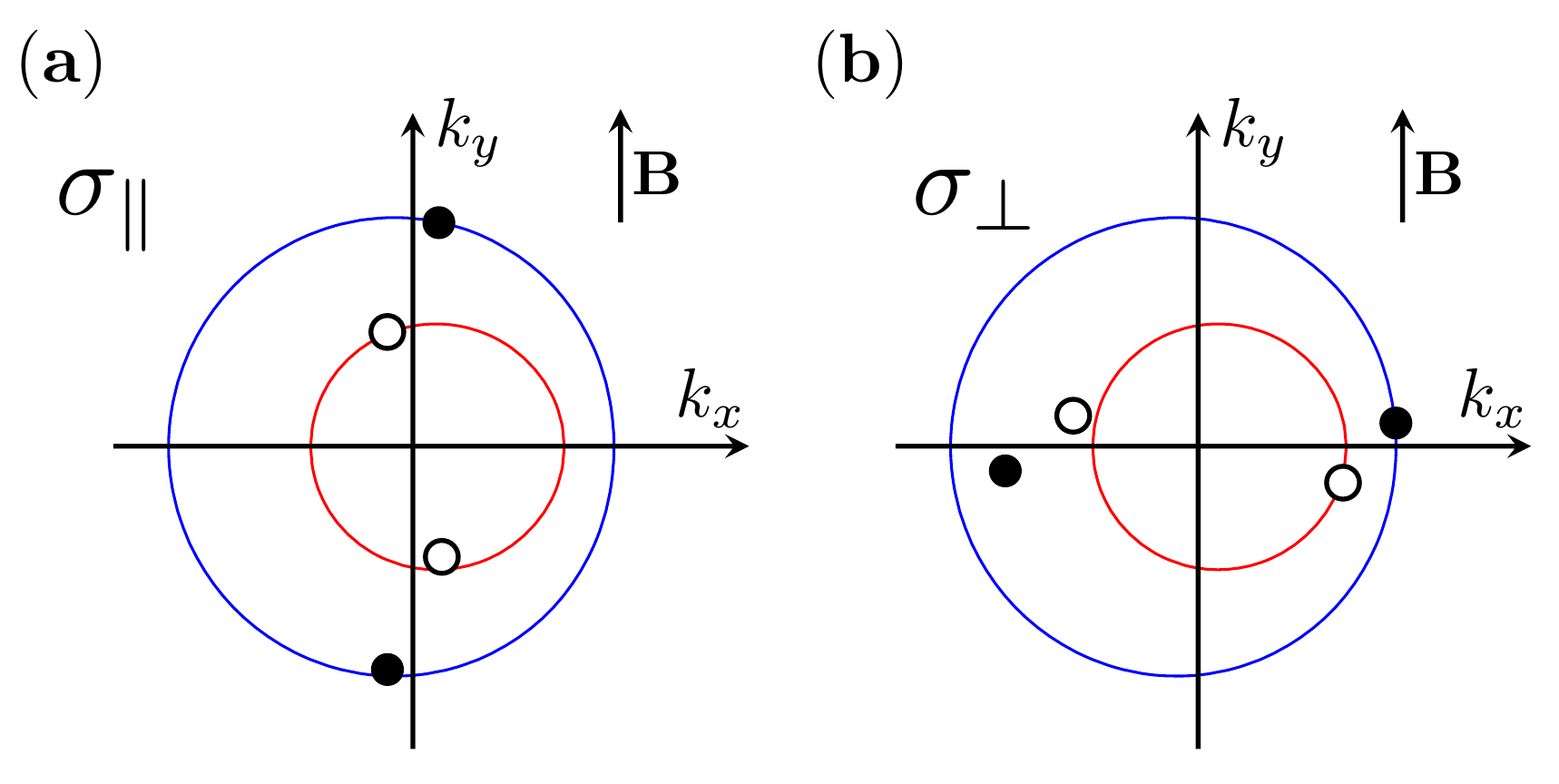}
    \caption{
        The in-plane field shifts the spin-split Fermi surfaces in the opposite directions. The Cooper pairs residing at the inner (outer) Fermi surfaces are shown as empty (full) circles.
        The typical Cooper pairs contributing to $\sigma_{\parallel}$ ($\sigma_{\perp}$) are shown in panel (a) [(b)].
        The pair breaking effect of the field is evident as not all the Cooper pairs can be placed at the Fermi level.
        The Cooper pairs shown in panel (a) experience a field induced pair breaking that is weaker than that experienced by the pairs shown in panel (b), implying $\sigma_{\parallel} > \sigma_{\perp}$.
    }
    \label{fig:PHE_pairs}
\end{figure}

The momentum and field dependence of $\varepsilon(\mathbf{q})$ is determined by the symmetries of the system. 
Irrespective of the crystal structure, time-reversal symmetry dictates that terms linear (quadratic) in $\mathbf{q}$ are odd (even) in $\mathbf{B}$. 
Additional constraints are contingent upon the point group symmetry of the crystal under consideration.

At the phenomenological level, the discussion remains general. 
As specific examples, we consider 2D superconductors with Rashba ($C_{3v}$ symmetry), Ising \cite{Lu2015,Xi2016,DeLaBarrera2018,Costanzo2018} ($D_{3h}$ symmetry), and Dresselhaus \cite{Winkler2003} ($D_{2d}$ symmetry) spin-orbit coupling. 
We demonstrate, based on symmetry arguments, that the PHE is finite (zero) for systems possessing (lacking) the $\sigma_h$ symmetry.

The most general form of the pair dispersion, encompassing Rashba and Ising symmetries, is given by:
\begin{align}\label{eq:dispersion}
     \varepsilon({\mathbf{q}}) = & \epsilon  + \xi^2 q^2 \!+\!  \mathcal{L}_1 (\mathbf{q} \times \mathbf{B}  )\cdot \hat{z}  
    \notag \\
     &    - \mathcal{L}_2 
    \left[(\mathbf{q}\cdot \mathbf{B})^2 - (\mathbf{q} \times \mathbf{B})^2 \right],
\end{align}
where $\xi^2$ represents the Cooper pair size averaged over all $\mathbf{q}$ directions, and it is minimally affected by a weak $\mathbf{B}$ field. 
Substituting Eq.~\eqref{eq:dispersion} into the general Eq.~\eqref{eq:conductivity} yields Eq.~\eqref{eq:res}.

Before delving into how symmetries govern the form of Eq.~\eqref{eq:dispersion}, let us first explore its physical meaning and its relation to Eq.~\eqref{eq:res}. 
To start, we can eliminate the $\mathcal{L}_1$ term by performing a shift of integration variables in Eq.~\eqref{eq:conductivity}. Consequently, this term does not contribute to the final result \eqref{eq:res}. 
Now, considering pairs propagating parallel ($\mathbf{q} \parallel \mathbf{B}$) or perpendicular ($\mathbf{q} \perp \mathbf{B}$) to the applied field, the dispersion relation \eqref{eq:dispersion} yields:
\(
    \varepsilon_{\parallel,\perp}(q) = \epsilon + (\xi^2 \mp \mathcal{L}_2) q^2
\),
where $\mp$ corresponds to parallel ($-$) and perpendicular ($+$) propagation, respectively.
For instance, if $\mathcal{L}_2 > 0$, it implies that Cooper pairs moving parallel to the applied field are bound more strongly (or have a smaller size) compared to those moving perpendicular to it.

To understand this anisotropy, let us consider the pair-breaking effect of the in-plane field in the Rashba model, as illustrated in Fig~\ref{fig:PHE_pairs}. In this model, Cooper pairs located at different regions of the Fermi surface are influenced in varying degrees by the applied field. Specifically, the in-plane field $\mathbf{B}$ causes shifts in the Fermi surfaces perpendicular to $\mathbf{B}$ in opposite directions \cite{Barnes2014}. This implies that pair breaking is weakest (strongest) for paired electrons propagating along (perpendicular to) $\mathbf{B}$, as depicted in Fig.~\ref{fig:PHE_pairs}(a) and Fig.~\ref{fig:PHE_pairs}(b), respectively.

Based on these geometric considerations, we anticipate that $\mathcal{L}_2 >0$ in the Rashba model. This expectation is consistent with the positive PHE described by Eq.~\eqref{eq:res}. 
Specifically, the supercurrent along the field is carried by pairs that are more strongly bound compared to those perpendicular to the field. 
This difference in binding strength contributes to the positive PHE.

%Now, let's return to the discussion of how symmetries determine the form of Eq.~\ref{eq:dispersion}, see Tab.~\ref{tab:symmetry}.
%The structure of the pair dispersion \eqref{eq:dispersion} is independent on the microscopic details.
%Instead, it is controlled by symmetries of the problem which we now turn to discuss.
%The symmetry constrains imposed on the $\mathcal{L}_{1,2}$ phenomenological constants are summarized in the 
%Tab.~\ref{tab:symmetry}.
%We now turn to the more detailed discussion of Eq.~\eqref{eq:dispersion} and Tab.~\ref{tab:symmetry}.
%%%% TABLE 1 %%%%%%%%

\begin{table}%[t!]
%\centering
\begin{tabular}{c||c|c|} 
 \hline
  2D ($\mathcal{P}\text{\sffamily X}$) & $\sigma_h \checkmark (D_{3h}) $ & $\sigma_h \text{\sffamily X} (C_{3v})$ \\  
 \hline
 \hline
 $\mathcal{L}_1$ & $= 0$ & $\neq 0$   \\
 \hline
 PHE, $\mathcal{L}_2$ & $ = 0$ & $ \neq 0$   \\
 \hline
\end{tabular}\, 
\caption{
Allowed ($\neq 0$) and forbidden ($0$) phenomenological constants $\mathcal{L}_{1}$ (second row) and  $\mathcal{L}_{2}$ (third row) in the pair dispersion, Eq.~\ref{eq:dispersion}.
The columns refer to a 2D systems with and without $\sigma_h$ symmetry. 
%two types of 2D systems discussed in Sec.~\ref{sec:phenomenology}.
The $C_{3v}$, (Rashba) and $D_{3h}$ (Ising) are representatives of these two classes of systems.
In all cases the parity $\mathcal{P}$ is broken.
}
\label{tab:symmetry}
\end{table}
%%%%%%%%%%

Now, let us return to the discussion of how symmetries determine the form of Eq.~\eqref{eq:dispersion}, as outlined in Table~\ref{tab:symmetry}. 
The term linear in momentum $\mathbf{q}$, proportional to the constant $\mathcal{L}_1$, is known as the Lifshitz invariant and has been extensively studied and tabulated (see, for instance, \cite{Smidman2017}). Specifically, the form of the Lifshitz invariant appearing in Eq.~\eqref{eq:dispersion} corresponds to the one allowed for the 2D Rashba system.

However, in systems where the $\sigma_h$ symmetry, such as the Ising superconductor, is present, the Lifshitz invariant is forbidden as it violates the combined $\sigma_h \mathcal{T}$ symmetry \cite{Fischer2018}. More generally, in the planar configuration, the breaking of $\sigma_h$ symmetry is a necessary condition for the Ginzburg-Landau free energy to contain terms that are odd in momentum.

Next, we examine the last term of the pair dispersion given by Eq.~\eqref{eq:dispersion}, which is proportional to $\mathcal{L}_2$.
To construct the symmetry-allowed terms note that the two pairs $(q_x,q_y)$ and $(B_y, - B_x)$ transform as identical two-dimensional $E$ irreducible representation of the $C_{3v}$ group \cite{Lax1974}.
Consequently, the only second-order combinations that transform trivially under $C_{3v}$ are $q^2$ and $B^2$. 
Additionally, we have two pairs of identically transforming combinations, $(2 q_x q_y, q_x^2- q_y^2)$ and $(2 B_x B_y, B_x^2- B_y^2)$, both belonging to the $E$ representation.
With this observation, we construct the scalar $4 q_x q_y B_x B_y + (q_x^2- q_y^2)(B_x^2- B_y^2) = (\mathbf{q}\cdot \mathbf{B})^2 - (\mathbf{q}\times \mathbf{B})^2$. 
The remaining scalar combination $q^2 B^2$ is fully isotropic and is included as part of the second term of Eq.~\eqref{eq:dispersion}.

To extend our analysis to the Dresselhaus symmetry, we begin with the two pairs $(q_x,q_y)$ and $(B_x,-B_y)$, which transform identically and irreducibly.
The resulting Lifshitz invariant, $q_x B_x - q_y B_y$, is well-documented for the $D_{2d}$ symmetry.
For the PHE, we need to identify the possible invariants that are second-order in both $q$ and $B$.
In addition to $q^2 B^2$ and $(\mathbf{q}\cdot \mathbf{B})^2 - (\mathbf{q}\times \mathbf{B})^2$, obtained for the Rashba symmetry case, we also have the combination $(q_x B_x - q_y B_y)^2$.
However, from the perspective of the PHE, this distinction is inconsequential, and both systems exhibit qualitatively similar field-induced anisotropy.

The symmetry group of the 2D Ising superconductor, $D_{3h}$, is derived from the Rashba symmetry group, $C_{3v}$, by adding $\sigma_h$.
In contrast to the Lifshitz invariant, the term proportional to $\mathcal{L}_2$ is quadratic in both momentum $\mathbf{q}$ and magnetic field $\mathbf{B}$.
Initially, it may seem that the addition of $\sigma_h$ symmetry does not influence the determination of the $\mathcal{L}_2$ constant.
Indeed, the three combinations, $q^2 B^2$, $(\mathbf{q} \cdot \mathbf{B})^2$, and $(\mathbf{q} \times \mathbf{B})^2 = q^2 B^2 - (\mathbf{q} \cdot \mathbf{B})^2$, are all permissible scalars for both $D_{3h}$ and $C_{3v}$ symmetries.
However, as we demonstrate, the $\sigma_h$ symmetry renders the Planar Hall effect zero in Ising superconductors.

The $\sigma_h$ symmetry dictates that the spin polarization of electrons, induced by the spin-orbit interaction points out-of-plane \cite{DeLaBarrera2018}. 
Thus, in addition to the crystallographic symmetries acting on both spin and orbital degrees of freedom, we have an additional symmetry $U_z(\delta \varphi)=\exp(- i \sigma_z \delta \varphi/2)$ that acts solely on spins. This operation rotates the spinors by an arbitrary angle $\delta \varphi$ around the $z$-axis.

In the planar configuration, $\mathbf{B}$ couples only to spins via a Zeeman interaction, $H_Z \propto \mathbf{B}\cdot \boldsymbol{\sigma}$. 
Consequently, Hamiltonians with differently oriented in-plane $\mathbf{B}$ are related by the unitary transformation, $U_z(\delta \varphi)$ for the angle $\delta \varphi$ between the two magnetic fields. 
This unitary equivalence implies that the system is fully isotropic with respect to the orientation of an in-plane $\mathbf{B}$. Consequently, $\mathcal{L}_{1,2} =0$, and the PHE is forbidden.

The specific expressions for $\mathcal{L}_2$ and its dependence on parameters such as spin-orbit splitting, electron mean free path and temperature can only be provided within a particular microscopic model, which is outlined in Sec.~\ref{sec:micro_model} for the Rashba spin-orbit interaction.
For readers uninterested in technical details, a summary of the analytical expressions for $\mathcal{L}_{1,2}$ is provided in Sec.~\ref{sec:summary_disp}.

\section{Microscopic Model}
\label{sec:micro_model}

We consider the two-dimensional disordered superconductor placed in an in-plane magnetic field $\mathbf{B}$, and described by the Hamiltonian, 
\begin{align}\label{eq:H}
    H = H_0 + H_{\text{SO}} + H_Z + H_{\text{dis}} + H_p\, , 
\end{align}
where $H_0$ is the kinetic energy of the free electrons, \( H_{\text{SO}} \) stands for the spin-orbit coupling present in a system without an inversion center, $H_Z$ is the Zeeman interaction term, $H_{\text{dis}}$ is the non-magnetic disorder potential, and $H_p$ is the pairing interaction.
Close to the $\Gamma$ point we can assume that the system has a $C_{\infty v}$ symmetry. 
In this limit the dispersion relation is parabolic with an effective mass $m$, 
\begin{align}\label{eq:H_0}
 H_0 = \sum_{\mathbf{k} s} c^{\dagger}_{\mathbf{k} s} (k^2 / 2 m) c_{\mathbf{k} s} \, ,     
\end{align}
where $c^{\dagger}_{\mathbf{k} s}$ creates an electron with momentum $\mathbf{k}$ and spin projection $s = \pm \tfrac{1}{2}$ on the \(z\) direction perpendicular to the basal $xy$-plane. 

The spin-orbit coupling takes the standard form, 
\begin{align}\label{eq:H_SO}
	H_{\text{SO}} 
		= \sum_{ \mathbf{k},ss' }  
		 c^{\dagger}_{\mathbf{k} s} 
		\left[ \boldsymbol{\gamma}(\mathbf{k}) \cdot  \boldsymbol{\sigma}\right]_{ s s' } c_{\mathbf{k} s'} \, ,
\end{align}
where $\boldsymbol{\sigma} = (\sigma_x,\sigma_y,\sigma_z)$ is the vector of Pauli matrices acting in spin space.
The $C_{3v}$ symmetry gives rise to Rashba spin-orbit coupling, 
$\boldsymbol{\gamma}(\mathbf{k}) = \alpha \left( \mathbf{k} \times \hat{z} \right) $.

The Zeeman coupling has a standard form, 
\begin{align}\label{eq:H_Z}
    H_Z = \sum_{ \mathbf{k},ss' }  
		 c^{\dagger}_{\mathbf{k} s} 
		\left[ \mathbf{B} \cdot \boldsymbol{\sigma} \right]_{ s s' } c_{\mathbf{k} s'} \, .
\end{align}

The spin-interactions \eqref{eq:H_SO} and \eqref{eq:H_Z} give rise to the two spin-split bands labeled by the index $\lambda = \pm $. Their energies counted relative to the Fermi energy $E_F$ is 
\begin{align}\label{eq:xiB}
	\xi^\lambda_{\mathbf{k},\mathbf{B}} = k^2/ 2m  + \lambda \alpha \tilde{k} - E_F \, ,
\end{align}
where $\tilde{\mathbf{k}} =  \mathbf{k} + \alpha^{-1} \hat{z} \times  \mathbf{B} $.
The spinors $\psi_\mathbf{B}^{\lambda}$ that make \(H_0 + H_{\text{SO}} + H_Z \) diagonal read as
\begin{align}\label{eq:psiB}
	\psi_\mathbf{B}^{\lambda} (\mathbf{k}) = \frac{1}{\sqrt{2}} 
	\begin{bmatrix}
         1 \\ \lambda \frac{ \tilde{k}_y - i \tilde{k}_x}{ \tilde{k} }
 \end{bmatrix} \, .
\end{align}
The inverse transformation reads as
\begin{align}\label{eq:psiB_inv}
    c_{\mathbf{k}\uparrow} & = 2^{-1/2} 
    \left[\psi_\mathbf{B}^{+} (\mathbf{k}) + \psi_\mathbf{B}^{-} (\mathbf{k}) \right]
    \notag \\
     c_{\mathbf{k}\downarrow } & = 2^{-1/2} 
    \left[\psi_\mathbf{B}^{+} (\mathbf{k}) - \psi_\mathbf{B}^{-} (\mathbf{k}) \right]
    (\tilde{k}_y + i \tilde{k}_x) \tilde{k}^{-1}\, .
\end{align}

In the limit $B=0$, we introduce the spinors of the chiral basis
\begin{align}\label{eq:chiral_basis}
    \psi_{\mathbf{k}}^\lambda  = \psi_{\mathbf{B}=0}^\lambda(\mathbf{k})\, .
\end{align}
The chiral basis spinors \eqref{eq:chiral_basis},  along with the zero-field dispersion (\( \xi^\lambda_{\mathbf{k}} = \xi^\lambda_{\mathbf{k},\mathbf{B}=0}\)) are obtained from Eqs.~\eqref{eq:psiB} and \eqref{eq:xiB} respectively by setting $\tilde{\mathbf{k}} = \mathbf{k}$.
The difference between the energy of the two chirality bands defines the spin-orbit energy splitting as 
\begin{align}\label{eq:DeltaSO}
    \Delta_{\mathrm{SO}} = \xi^+_{\mathbf{k}_F}  - \xi^-_{\mathbf{k}_F} = 2 \alpha k_F\, .
\end{align}
At $B=0$, the two spin split Fermi momenta are \( k^\lambda_F = \sqrt{\alpha^2 m^2 + k_F^2} - \lambda m \alpha \) with \( k_F = \sqrt{2 m E_F}\). 
The Fermi velocity, $v_{F} = \sqrt{v_{F0}^2 + \alpha^2}$, $v_{F0} = k_F/m$ is the same for both chiralities $\lambda = \pm 1$.

The density of states for the two bands reads $\nu_\lambda = \nu_0 (1 - \lambda x)$, where $x =\Delta_{\mathrm{SO}} /4 E_F$.
The finite difference of the two densities of states, $\nu_- - \nu_+ = 2 \nu_0 x$ is necessary 
for the Lifshitz invariant in the limit of weak magnetic field. 
One of our results is that this is not strictly speaking true once the Cooper pairing in the triplet channel is taken into account, repulsive or attractive alike. 
Hence we turn to the description of the pairing interaction with this observation in mind.

For simplicity we consider the short range spin conserving disorder potential of the form,
\begin{align}\label{eq:Hdis}
    H_{\text{dis}} = V\sum_{\mathbf{R}_j} \sum_{\mathbf{k}',\mathbf{k},s} e^{i (\mathbf{k}'- \mathbf{k})\cdot \mathbf{R}_j} c^\dag_{\mathbf{k}s}c_{\mathbf{k}'s}\, ,
\end{align}
where the summation over the scattering centers labeled by $j$ and placed at random locations, $\mathbf{R}_j$.
The disorder, Eq.~\eqref{eq:Hdis}, gives rise to the disorder scattering rate,
\(1/\tau = 2 \pi \nu_0 |V|^2\).

The pairing interaction normally contains singlet and triplet parts,
$H_p = H_{p}^s + H_{p}^t$. 
The triplet interaction is presented in Appendix~\ref{app:pairing_t}.
The singlet part is standard 
\begin{align}\label{eq:Hps}
    H_{p}^s = &  \frac{g}{4} \sum_{\mathbf{k},\mathbf{k}',\mathbf{q};s} 
		\left\{ c^\dagger_{\mathbf{k}_+s_1} \left[ i \sigma_y \right]_{s_1,s_2} c^\dagger_{-\mathbf{k}_- s_2} \right\}
  \notag \\
  & \times  \left\{ c_{-\mathbf{k}'_-s_3} \left[ i \sigma_y \right]^\dagger_{s_3 s_4} c_{\mathbf{k}'_+s_4} \right\}\, ,
\end{align}
where we have introduced the notation $\mathbf{k}_{\pm} =\mathbf{k} \pm \mathbf{q}/2$, and $\sum_{\mathbf{k},\mathbf{k}',\mathbf{q};s}$ denotes the summation over the momenta $\mathbf{k}$, $\mathbf{k}'$ and $\mathbf{q}$ as well as over all the spin indices. 
The coupling $g$ fixes the critical temperature, 
$T_{c0} = (2 e^{ \gamma_E } /\pi ) \omega_D \exp(- 1/ |g| \nu_0 )$, where $\gamma_E$ is the Euler gamma constant, and $\omega_D$ is the Debye frequency.

\begin{figure}
    \centering
    \includegraphics[scale = 0.4]{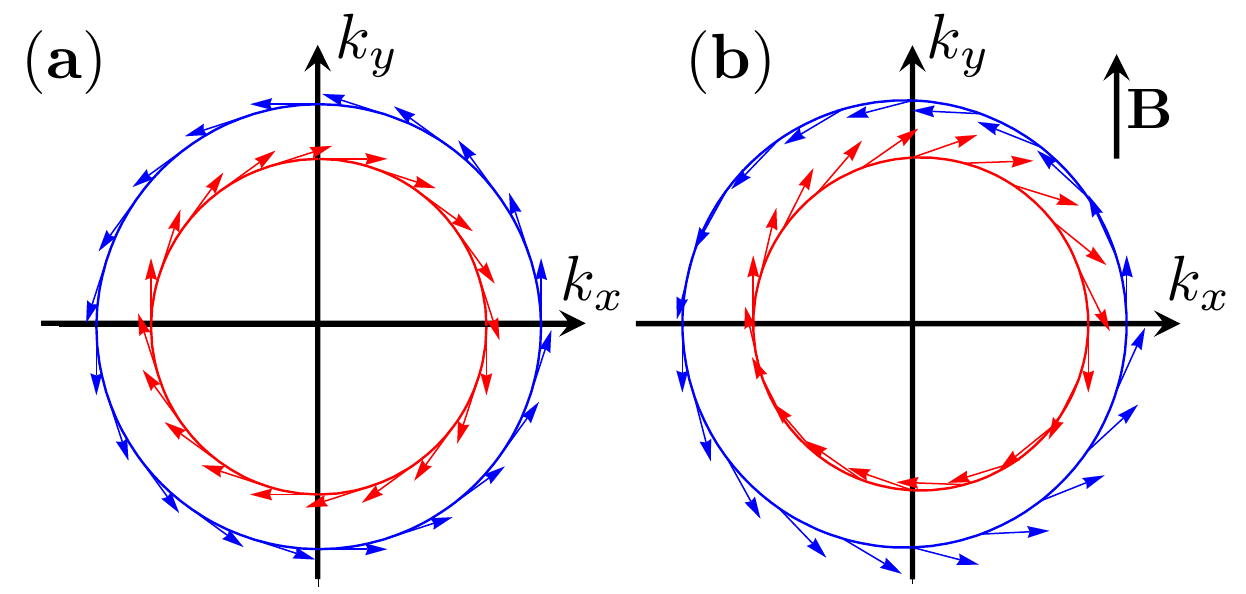}
    \caption{Deformation of the spin texture set by Rashba spin-orbit interaction due to the in-plane field (arbitrary units).
    The two circles are the Fermi surfaces of the two spin split bands.
    The arrows stand for the spin polarization at the point of origin. 
    (a) 
    We set $\Delta_{\mathrm{SO}} = 0.6 E_F$ for illustration. 
    (b) An in-plane field $\mathbf{B}= B \hat{y}$ shifts the two Fermi surfaces in opposite directions. The deformation of the spin textures is largest along the $k_y$ axis, where the combination $\boldsymbol{\gamma}(\mathbf{k}) \times \mathbf{B}$ is maximal.
    }
    \label{fig:texture}
\end{figure}

\subsection{The regime of parameters to compute \texorpdfstring{\(\mathcal{L}_{2}\)}{ℒ2} }

We now formulate the regime for which the calculation presented in Sec.~\ref{sec:calculation} holds.
We have assumed that the Zeeman energy is sufficiently small,
$B \ll \min \{ T_c, 1/\tau \}$.
Here $1/\tau$ is the disorder scattering rate. 
In Eq.~\eqref{eq:res} and below we have set the $g$ -factor to two,  
and have absorbed the Bohr magneton $\mu_B$ into the definition of $\mathbf{B}$ for clarity.
The parameters specifying the pair dispersion \eqref{eq:dispersion} summarized in Sec.~\ref{sec:summary_disp} for the clean and dirty limits, $T_c \gg 1/\tau$ and $T_c \ll 1/\tau$, respectively.
The typical dependence of the conductivity [Eq.~\eqref{eq:res}] on temperature and disorder scattering rate is shown in Fig.~\ref{fig:sigmaXY}.

In all of the calculations we assume that the spin-orbit energy splitting $\Delta_{\mathrm{SO}}$ at the Fermi level is larger than other energy scales except for the Fermi energy, $E_F$.
The parameter range we have specifically considered is,
\begin{align}\label{eq:range}
  B \ll \min\{ T_c, 1/\tau \}\, , \max \{ T_c, 1/\tau \}  \ll  \Delta_{\mathrm{SO}} \ll E_F\, .
\end{align}

The second condition in Eq.~\eqref{eq:range} disfavours the inter-band Cooper pairs, therefore essentially implying that the two spin split bands can be treated as a two-band superconductor.
The extension of the present treatment beyond the limit set by Eq.~\eqref{eq:range} is relegated to future studies. 
In the considered limit of the effective two-band superconductor the quasi-classical theory has been previously developed, \cite{Houzet2015}.
We have checked the consistency of our approach and the quasi-classical theory reproducing the earlier results for $\mathcal{L}_1$ in special cases.

Below in Sec.~\ref{sec:calculation} we provide all the details of the calculation of the dispersion relation~\eqref{eq:dispersion}.
For convenience, prior to delving into the details, we list the coefficients entering the dispersion relations in the clean and dirty limits in Sec.~\ref{sec:summary_disp}.

\subsection{Summary of the dispersion relation \texorpdfstring{Eq.~\eqref{eq:dispersion}}{Eq.3}}
\label{sec:summary_disp}

We discuss now in turn all terms in the dispersion relation \eqref{eq:dispersion} with different order in momentum and field. 
We give the general result and the expressions in the clean and dirty limits.

\subsubsection{Critical temperature suppression}
To zero order in both momentum and magnetic field, Eq.~\eqref{eq:dispersion} describes the critical temperature reduction caused by the applied field,
\begin{align}\label{eq:TcR}
    T_c - T_{c0} = - 4 T_{c0} B^2  \tau \gamma_2 \,,
\end{align}
where \( T_{c0} = T_c(B=0) \). The suppression of the critical temperature is detailed in Appendix~\ref{app:polarization_bubble}.
Here we have introduced the useful sums over the Matsubara frequencies, $\varepsilon_n = 2 \pi T ( n+1/2)$,
\begin{subequations}\label{eq:notation_sum}
\begin{align}\label{gamma_def}
\gamma_j 
	& =  T \sum_{ \varepsilon_n>0 } \frac{\pi}{ \varepsilon_n^2 \left( 1 + 2 j \tau \varepsilon_n \right) } \, ,
 \\ \label{eq:u_k}
	u_j
	& = T \sum_{ \varepsilon_n >0 } \frac{\pi}{ 2 \varepsilon_n^j \left( 1 + 2 \tau \varepsilon_n \right)^3 \left( 1 + 4 \tau \varepsilon_n \right)^2 } \, ,
\end{align}
\end{subequations}
where \(j\) is non-negative integer. The origin of these definitions is detailed in Appendix~\ref{app:polarization_bubble}.

In the clean and dirty limit $\gamma_j$, Eq.~\eqref{gamma_def} attains the values 
\begin{align}\label{eq:gamma_limits}
   \tau \gamma^c_j = \frac{2}{k} \frac{ \left( \xi^c_0 \right)^2}{v_F^2} \, ,\,\,\,
   \tau \gamma^d_j = 2 \frac{ \left( \xi^d_0 \right)^2}{v_F^2} \,,
\end{align}
expressible in terms of the zero-field size of the Cooper pairs $\xi_0$ in their respective limits:
\begin{align}\label{eq:xi_0_limits}
    \xi_0^c  = \left( \frac{7 \zeta(3) v_F^2}{ 32 \pi^2 T_c^2 }\right)^{1/2}, \,  \,
    \xi_0^d  = \left( \frac{ \pi \tau v_F^2 }{ 16 T_c } \right)^{1/2}\, .
\end{align}
Substituting the clean limit $\tau \gamma^c_2$ given by Eq.~\eqref{eq:gamma_limits} into Eq.~\eqref{eq:TcR} reproduces the result of Ref.~\cite{Barzykin2002}.
In the dirty limit we obtain instead \( T_c - T_{c0} = -   B^2 \pi \tau / 2\).
Comparison of the clean and dirty limits shows that disorder opposes the pair breaking effect of the in-plane field. 
This is in contrast to the case of Ising superconductors with $\boldsymbol{\gamma}(\mathbf{k}) \parallel \hat{z}$, \cite{Bulaevskii976,Frigeri2004,Ilic2017}.

The superconducting transition occurs into the helical state with finite momentum, 
\(\mathbf{q}_h = \hat{z} \times \mathbf{B} \mathcal{L}_1/ 2 \xi_0^2 \).
This, however has negligible effect on the critical temperature.

\subsubsection{Lifshitz invariant \texorpdfstring{\(\mathcal{L}_{1}\)}{ℒ1}}
As we stressed in Sec.~\ref{sec:phenomenology}, the Lifshitz invariant bears no implications on PHE.
To maintain consistency the terms of the pair dispersion that are of first order in momentum have to be analyzed before we come to the second order terms.  
The calculation of Lifshitz invariant is a benchmark for the subsequent analysis of $\mathcal{L}_2$.
We recover the previous results of Refs.~\cite{Houzet2015,Edelstein2021} for the Lifshitz invariant.
Yet, when the Cooper channel coupling is not too weak we find additional contributions.

The standard result for $\mathcal{L}_1$ reads,
\begin{align}\label{eq:L1}
    \mathcal{L}_1 = x   v_F \tau \gamma_2
\end{align}
(see Appendix~\ref{app:polarization_bubble} and \ref{app:L} for the detailed derivation).
The clean and dirty limits of Eq.~\eqref{eq:L1} follow from Eqs.~\eqref{eq:gamma_limits} and \eqref{eq:xi_0_limits}.
The suppression of $\mathcal{L}_1$ caused by disorder follows from the ratio of $\mathcal{L}^{c,d}_1$ in the clean and dirty limits,
\begin{align}\label{eq:L1_ratio}
    \frac{\mathcal{L}^d_1 }{\mathcal{L}^c_1} = \frac{2 \pi^3}{7 \zeta(3)} T_c \tau\, .
\end{align}

We stress that a finite result is obtained only if the density of states of the two spin split bands are not the same, i.e. $x\neq 0$.
The dependence of $\mathcal{L}_1$ on the disorder strength and the temperature are the same as that of the $T_c$ suppression as both are proportional to $\tau \gamma_2$.
Thus stronger pair breaking leads to enhanced $\mathcal{L}_1$.

Apart from Eq.~\eqref{eq:L1} there is an additional contribution to $\mathcal{L}_1$,
\begin{align}\label{eq:deltaL1}
    \delta \mathcal{L}^s_1 =
        \ln( \frac{ 2 \omega_D }{ \pi T_{c0} } e^{ \gamma_E }) 
        \frac{1}{ k_F \Delta_{\mathrm{SO}}}
\end{align}
originating from the modification of the matrix elements of the spin singlet interaction, Eq.~\eqref{eq:Hps} by the field, see Appendix~\ref{app:pairing_s} for details.
This can be interpreted as arising from the field induced deformation of the spin texture, see Fig.~\ref{fig:texture}.
Here we underestimate $\mathcal{L}^s_1$ by ignoring $\log(\omega_D/\Delta_{\mathrm{SO}})$ in comparison to $\log(\omega_D/T_{c0})$. 
Otherwise Eq.~\eqref{eq:deltaL1} acquires an additional factor of 2.

At first glance in the considered range of parameters [Eq.~\eqref{eq:range}], Eq.~\eqref{eq:deltaL1} is smaller than the standard expression for the Lifshitz invariant,  Eq.~\eqref{eq:L1}, which is proportional to $T_c^{-1}$, Eq.~\eqref{eq:xi_0_limits}.
Nevertheless, Eq.~\eqref{eq:deltaL1}  can be comparable or even exceed Eq.~\eqref{eq:L1} because it is enhanced by the large Cooper logarithm. 
Furthermore, Eq.~\eqref{eq:L1} is proportional to $x \ll 1$.
Specifically, in the clean limit, the ratio
\begin{align}\label{eq:ratioL1c}
    \frac{\delta \mathcal{L}_1^s}{\mathcal{L}_1^c} = 
     \ln( \frac{ 2 \omega_D }{ \pi T_{c0} } e^{ \gamma_E }) \frac{T_c^2}{ \Delta^2_{\mathrm{SO}}} 
      \frac{ (8 \pi)^2  }{7 \zeta(3) }
\end{align}
can be large within the considered range of parameters, Eq.~\eqref{eq:range}.
For instance, taking $\omega_D = 250$K, $\Delta_\mathrm{SO} = 50$K in Eq.~\eqref{eq:ratioL1c} gives $\delta \mathcal{L}^s_1 / \mathcal{L}_1^c \approx 3.2$ for $\abs{g} \nu_0 = 0.25$.
At the same time $T_{c0} \approx 5$K still satisfies the condition, $T_{c0} \ll \Delta_\mathrm{SO}$.
The ratio Eq.~\eqref{eq:ratioL1c} is enhanced by disorder. 
For sufficiently weak coupling the exponential decrease of $T_c$ makes this ratio inevitably small, and the standard result, Eq.~\eqref{eq:L1}, is recovered.

We note in passing that the $p$-wave triplet channel interaction, Eq.~\eqref{eq:Hpt}, makes a contribution to the Lifshitz invariant that is similar in form to that of the interaction in the singlet channel, Eq.~\eqref{eq:deltaL1},
\begin{align}\label{eq:deltaL1_triplet}
    \delta \mathcal{L}^t_1 = - 4 \frac{g_t}{|g|} \delta \mathcal{L}^s_1 \, ,
\end{align}
where $g_t$ is the triplet interaction amplitude that is assumed to be weak, \(g_t \ll g\). 
Equation~\eqref{eq:deltaL1_triplet} is derived in Appendix~\ref{app:pairing_t}.

\subsubsection{Field dependence of the Cooper pair size, \texorpdfstring{\(\xi\)}{ξ} }
To second order in $B$ we write $\xi^2 = \xi_0^2 + \xi_2^2 B^2$, where up to a small corrections in $x \ll 1$
\begin{align}\label{eq:xi_0}
    \xi_{0}^2 =  \frac{1}{2}v_F^2 \tau \gamma_1\, 
\end{align}
is standard, \cite{Kopnin2001}.
In the clean and dirty limits, $\xi_{0}$ reduces to $\xi_{0}^{c,d}$ given by Eqs.~\eqref{eq:xi_0_limits} respectively. 

In a weak field, the angular average size of the Cooper pairs decreases by an amount 
\begin{align}\label{eq:xi_2}
    \xi_{2}^2 = - \rho_2 - \rho_1/4\, ,
\end{align}
where as in Eq.~\eqref{eq:xi_0} we have set $x$ to zero. Eq.~\eqref{eq:xi_2} is derived in Appendix~\ref{app:L}. We introduce the notation
\begin{subequations}\label{eq:rhos}
	\begin{align}
		\rho_1
		= &
		2 \pi \tau^2 v_F^2  
		\left( 13 \tau u_2 + 72 \tau^2 u_1 + 96 \tau^3 u_0 \right)\, ,
		\label{rho_1}
		\\
		\rho_2
		= &
		\frac{\pi}{2} \tau^2 v_F^2
		\left( 4 u_3 + 37 \tau u_2 + 104 \tau^2 u_1 + 96 \tau^3 u_0 \right)
		 \, ,
		\label{rho_2}
	\end{align}
\end{subequations}
where we employed the previous definition of \( u_k \), Eq.~\eqref{eq:u_k}.
The expression arise from the Matsubara summation of the bubble diagram (see Appendix~\ref{app:polarization_bubble}).

In the clean limit these are reduced to
\begin{align}\label{eq:rho_c}
    \rho^c_1 = 93 \zeta(5) \left( \frac{4}{7 \zeta(3)} \right)^2  \frac{ \left( \xi^c_0 \right)^4}{v_F^2}\, , \quad  \rho^c_2 =  \rho^c_1 /4\, .
\end{align}
and in the dirty limit, 
\begin{align}\label{eq:rho_d}
    \rho^d_1 = 52 \tau^2 \left( \xi_0^d \right)^2\, , \quad 
    \rho^d_2 = \frac{14 \zeta(3)}{ \pi^2} \frac{\tau}{T_c} \left( \xi_0^d \right)^2\, .
\end{align}

\subsubsection{The field induced anisotropy, \texorpdfstring{\(\mathcal{L}_{2}\)}{ℒ2}}

The field induced anisotropy of the Cooper pairs is encapsulated in the $\mathcal{L}_2$ coefficient,  
\begin{align}\label{eq:L2}
    \mathcal{L}_2 = \rho_1/4\, ,
\end{align}
(see Appendix~\ref{app:polarization_bubble} and \ref{app:L} for a more detailed derivation).
Based on Eqs.~\eqref{eq:rho_c} and \eqref{eq:rho_d} the ratio of the anisotropy coefficient in the clean and dirty limits is
\begin{align}\label{eq:L2_ratio}
    \frac{\mathcal{L}^d_2 }{\mathcal{L}^c_2 } = (\tau T_c)^3 \frac{206 \pi^5}{93 \zeta(5)}\, .
\end{align}
Equation~\eqref{eq:L2_ratio} indicates that disorder suppresses the planar Hall effect. 
This is to be expected as the disorder tends to restore the isotropy of the dispersion relation of the Cooper pairs.

Similar to the case of the Lifshitz invariant the variation of the matrix elements of the singlet pairing interaction as well as the $p$-wave triplet interaction introduce corrections to $\mathcal{L}_2$.
These corrections read as
\begin{align}\label{eq:deltaL2}
        \delta \mathcal{L}_2^s = \ln( \frac{ 2 \omega_D }{ \pi T_{c0} } e^{ \gamma_E }) \frac{1 }{ 2 k_F^2 \Delta_{\mathrm{SO}}^2}\, ,
\quad 
         \delta \mathcal{L}^t_2 = \frac{g_t}{|g|} \delta \mathcal{L}^s_2 \, ,    
\end{align}
respectively. These are derived in Appendix~\ref{app:pairing_high}. 
Unlike the case of $\mathcal{L}_1$ these corrections are small,
\begin{align}\label{eq:ratioL2c}
    \frac{\delta \mathcal{L}_2^{s,t}}{\mathcal{L}_2} \propto 
    \ln( \frac{ 2 \omega_D }{ \pi T_{c0} } e^{ \gamma_E }) \left( \frac{T_c}{E_F} \right)^2 \left( \frac{T_c}{\Delta_\mathrm{SO}} \right)^2\, . 
\end{align}
In contrast to Eq.~\eqref{eq:ratioL1c}, Eq.~\eqref{eq:ratioL2c} contains an extra small parameter $(T_c/E_F)^2$, which makes \( \delta \mathcal{L}^{s,t}_2\) irrelevant.
The difference between $\mathcal{L}_1$ and $\mathcal{L}_2$ stems from the different dependence on the density of states $\mathcal{L}_{1,2} \propto \nu_- \mp \nu_+$.

%The reason for this difference is that in contrast to $\mathcal{L}_2$, $\mathcal{L}_1$ hinges upon the finite difference of the density of states at the two spin split Fermi surfaces.
%This increases the relative importance of the corrections $\delta \mathcal{L}_1^{s,t}$.

\section{Calculation of the dispersion of the Cooper pairs}
\label{sec:calculation}

In this section we perform the calculation of the Cooper pair dispersion~\eqref{eq:dispersion}.
For now we focus on the pairing in the s-wave singlet channel.
At a latter stage we analyze the additional contributions due to the triplet channel interaction.
Introduce the correlation function for the Cooper pairs at the momentum $\mathbf{q}$ and bosonic Matsubara frequency, \( \Omega_n = 2\pi T n\) \cite{Larkin2005}:
\begin{align}\label{eq:L}
	K & \left( \mathbf{q}, i \Omega_m \right)
	=
	 \sum_{\mathbf{k},\mathbf{k}',s} \int_0^{T^{-1}} d \tau \exp(i \Omega_n \tau ) 
 \notag \\
	  \times &  \Big\langle
		T_\tau \big\{ c_{-\mathbf{k}_- s_1}(\tau) \left[ i \sigma_y \right]^\dagger_{s_1 s_2} 
 c_{\mathbf{k}_+ s_2}(\tau)
 \notag \\
& \times  c_{\mathbf{k}'_+ s_1 }^\dagger(\tau) \left[ i \sigma_y \right]_{s_1,s_2} c_{-\mathbf{k}'_-s_2 }^\dagger(\tau) \big\} 
  \Big\rangle\, ,
\end{align}
where $T_\tau$ stands for the time ordering in the imaginary time $\tau$, and for any operator $O$, 
\( O(\tau) = \exp(H \tau) O \exp( - H \tau) \).

The dynamic properties of the Cooper pairs are contained in the retarded correlation functions obtained from \eqref{eq:L} via the analytic continuation, $i \Omega_n \rightarrow \Omega + i 0_+$.
In this work we assume that the dependence of the Cooper pair propagators on the frequency $\Omega$ at $\Omega \lesssim T - T_c \ll T_c$ is unaffected by the magnetic field and the spin orbit interaction.
Under this condition we take the Cooper pairs dissipation rate $\gamma_{\text{GL}}$ in Eq.~\eqref{eq:conductivity} as in the standard BCS theory.
This assumption holds under the conditions \eqref{eq:range}.
We therefore set $\Omega_n =0$ in Eq.~\eqref{eq:L} above.

In the weak coupling regime, $|g| \nu_0 \ll 1$ the correlation function \eqref{eq:L} takes the standard form 
\begin{align}\label{eq:K}
    K \left( \mathbf{q}, i \Omega_m \right) = \frac{\Pi \left( \mathbf{q}, i \Omega_m \right)}{1 + (g/4)\Pi \left( \mathbf{q}, i \Omega_m \right) }\, ,
\end{align}
where the polarization operator, $\Pi \left( \mathbf{q} \right) $ at \( \Omega_m = 0 \), is the correlation function $K$ in the non-interacting limit, $g=0$.

The dispersion relation is proportional to the denominator of Eq.~\eqref{eq:K},
\begin{align}\label{eq:pair_propagator}
	\nu_0 \varepsilon \left( {\mathbf{q}} \right)  = 1/g + \Pi\left( {\mathbf{q}}\right) /4 \, ,
\end{align}
where as before the order parameter is normalized to equal the gap function at equilibrium.

\subsection{Chiral basis formulation}
\label{sec:chiral}
The calculation of the Cooper pair dispersion, \eqref{eq:pair_propagator} is performed in the chiral basis, Eq.~\eqref{eq:chiral_basis}. 
This is done in two steps.
First we employ Eq.~\eqref{eq:psiB} to transform to the basis which diagonalizes the free part of the Hamiltonian \eqref{eq:H} including Eqs.~\eqref{eq:H_0}, \eqref{eq:H_SO} and \eqref{eq:H_Z}.
Second, one performs the expansion of the dispersion relation \eqref{eq:pair_propagator} in $B$.
As $B$ is set to zero, the generic basis Eq.~\eqref{eq:psiB} turns into the chiral basis, Eq.~\eqref{eq:chiral_basis}.

The transformation to the basis, Eq.~\eqref{eq:psiB} has an advantage of making the Green function matrix diagonal, \( G_{\lambda \lambda'} = \delta_{\lambda\lambda'} G_{\lambda} \) even at finite $B$.
The diagonal elements of the Green function read as
\begin{align}\label{eq:G_lambda}
    G_{\lambda}(\mathbf{k},\varepsilon_n) = 
\left[
	i \varepsilon_n 
	- \xi^\lambda_{ k } - \mathbf{v}_{F} \cdot \mathbf{Q}^\lambda
	+ i \frac{ \mathrm{sgn}(\varepsilon_n) }{ 2 \tau }
	\right]^{-1} \, ,
\end{align}
where 
\begin{align}\label{eq:Q}
    \mathbf{Q}^\lambda = \mathbf{q}/2 + \lambda \hat{z}\times \mathbf{B}/v_F\, .
\end{align}
We note that the diagonal form of the Green function presumed by the Eq.~\eqref{eq:G_lambda} is preserved by the disorder potential only if the disorder potential does not mix spin split bands.
This imposes the condition \(\max\{1/\tau,T_c \} \ll \Delta_{\mathrm{SO}}\) included in Eq.~\eqref{eq:range}.

The calculations are performed to the fourth order in $\mathbf{Q}^\lambda$.
This implies according to the definition \eqref{eq:Q} that we keep all the terms in the Cooper pair dispersion to the order $n_q$ in \(q v_F/ \max\{T_c, 1/\tau\}\)  and $n_B$ in \(B/ \max\{T_c, 1/\tau\}\) such that $n_q+n_B \leq 4$.
Clearly, such procedure is sufficient to obtain all the terms in the expression \eqref{eq:dispersion}.
Each of the coefficients in Eq.~\eqref{eq:dispersion} is given in Sec.~\ref{sec:summary_disp} in clean and dirty limits.

\begin{figure}
    \centering
    \includegraphics[scale = 0.4]{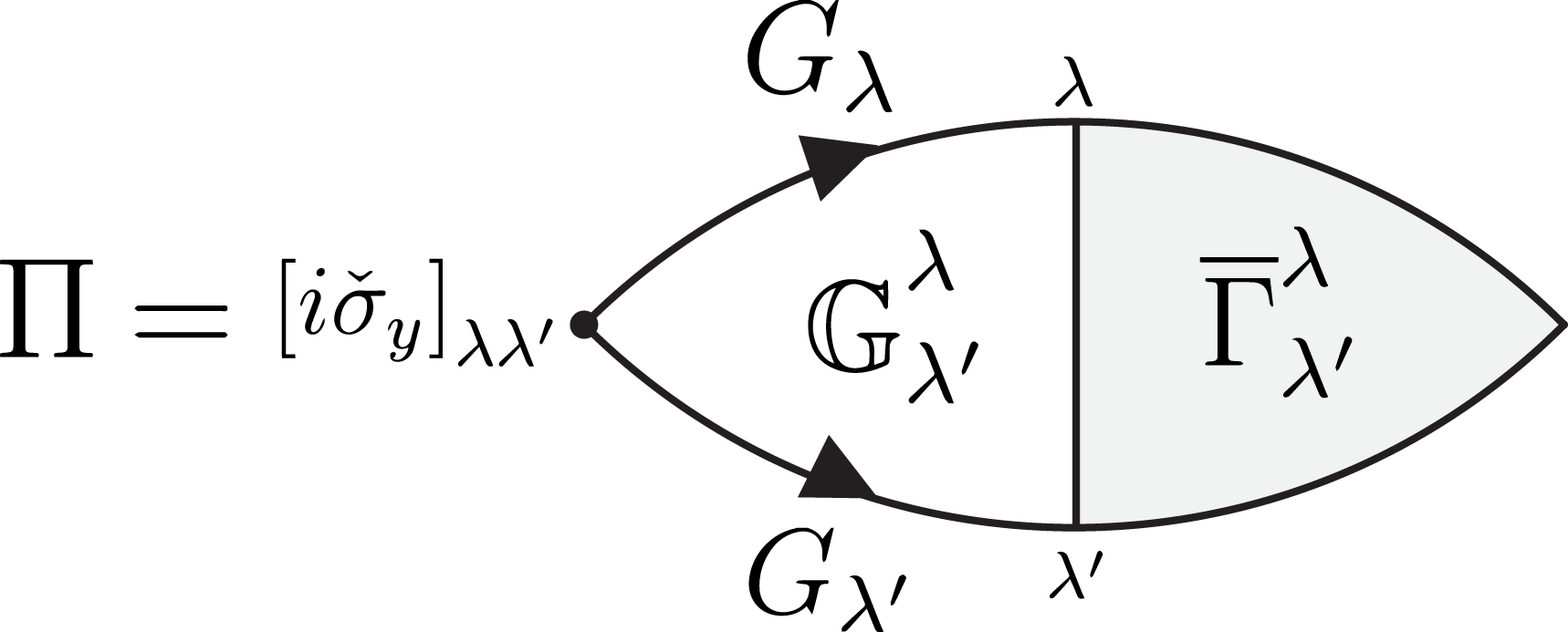}
    \caption{Diagrammatic presentation of the equation \eqref{eq:Bubble_w_Cooperon} defining the effect of the disorder on the fluctuation spectrum. 
    $\mathbb{G}^{ \lambda }_{ \lambda' }$ is introduced in Eq.~\eqref{eq:GF_pair}, and $\overline{\Gamma}^{ \lambda }_{ \lambda' }$ stands for the Cooperon vertex.
    }
    \label{fig:Pi}
\end{figure}

To encompass the effect of the disorder we express the Cooper pair propagator in terms of the Green functions and the Cooperon vertex, \(\overline{ \Gamma }^{ \lambda }_{ \lambda' }(\mathbf{k},\mathbf{q},\mathbf{B},\varepsilon_n)\), see Fig.~\ref{fig:Pi}, 
\begin{align}\label{eq:Bubble_w_Cooperon}
	\! \Pi \left( \mathbf{q} \right)  \!	 =\! 2 T\!\! 
	\sum_{\mathbf{k},\varepsilon_n \lambda \lambda'   }\!\!
	\left[ i  \check{\sigma}_y (\mathbf{k},\mathbf{q},\mathbf{B}) \right]_{ \lambda \lambda' }
	\mathbb{G}^{ \lambda }_{ \lambda' } \left( \mathbf{k}, \mathbf{q}, \mathbf{B}, \varepsilon_n \right)
	\overline{ \Gamma }^{ \lambda }_{ \lambda' }\, ,
\end{align}
where the combinatorial factor of 2 is included,
\begin{align}\label{eq:GF_pair}
	\mathbb{G}^{ \lambda }_{ \lambda' } \left( \mathbf{k}, \mathbf{q}, \mathbf{B}, \varepsilon_n \right)
         = 
         G_\lambda \left( \mathbf{k}_+,\varepsilon_n \right) G_{\lambda'} \left(- \mathbf{k}_-,-\varepsilon_n \right)\, ,
\end{align}
and $\left[ i \check{\sigma}_y (\mathbf{k},\mathbf{q},\mathbf{B}) \right]$ stands for the interaction vertex of the interaction Hamiltonian [Eq.~\eqref{eq:Hps}] in the basis of Eq.~\eqref{eq:psiB}.
It defined such that the interaction Hamiltonian, \eqref{eq:Hps} takes the form 
\begin{align}\label{eq:Hps_chiral}
    H_{p}^s 
        = &  \frac{g}{4} \sum_{\mathbf{k},\mathbf{k}',\mathbf{q};\lambda} 
		\left\{
            \hat{\psi}_{\mathbf{B}}^{\lambda_1\dagger}(\mathbf{k}_+) \left[ i \check{\sigma}_y \right]_{\lambda_1,\lambda_2} \hat{\psi}_{\mathbf{B}}^{\lambda_2 \dagger}(-\mathbf{k}_- ) \right\}
        \notag \\
            & \times  \left\{ \hat{\psi}_{\mathbf{B}}^{\lambda_3}(-\mathbf{k}'_-) \left[ i \check{\sigma}_y \right]^\dagger_{\lambda_3 \lambda_4} \hat{\psi}_{\mathbf{B}}^{\lambda_4}(\mathbf{k}'_+) 
        \right\}\, .
\end{align}
Naturally, \(\left[ i  \check{\sigma}_y (\mathbf{k},\mathbf{q},\mathbf{B}) \right]\) vertex takes a more complex form of Eq.~\eqref{SC_Pauli_chiral} than just $i \sigma_y$ in the original basis~\eqref{eq:Hps}.
We will see, however, that since the interaction vertex depends on the two small parameters,  $q/k_F \ll 1$ and $B/\Delta_{\mathrm{SO}} \ll 1$ it is greatly simplified in the studied limit.

In contrast  to the interaction vertex in Eq.~\eqref{eq:Bubble_w_Cooperon}, the second factor, \(\mathbb{G}^{ \lambda }_{ \lambda' }\) depends on the momentum and the field via the dimensionless parameters, \(q v_F/ \max\{T_c, 1/\tau\}\) and \(B/ \max\{T_c, 1/\tau\}\). 
Because as stated in Eq.~\eqref{eq:range}, \(\max\{T_c, 1/\tau\} \ll \Delta_{\mathrm{SO}}\) and \(T_c \ll E_F \) we can discard the momentum and the field dependence of 
$ i  \check{\sigma}_y (\mathbf{k},\mathbf{q},\mathbf{B}) $ and keep it only in the $\mathbb{G}^{ \lambda }_{ \lambda' }$.
As detailed in Appendix.~\ref{app:inter-band_zero_order}, to zero order in $q/k_F$ and $B/\Delta_{\mathrm{SO}}$, the interaction vertex,
\begin{align}\label{eq:check_sigma}
    \left[ i  \check{\sigma}_y (\mathbf{k}) \right]_{ \lambda \lambda' } 
		 = i \lambda \delta_{  \lambda \lambda' } e^{ i \varphi}\, 
\end{align}
is purely intra-band.

\subsection{Cooperon vertex renormalization}
\label{sec:Cooperon}

\begin{figure}
    \centering
    \includegraphics[width = 0.48\textwidth]{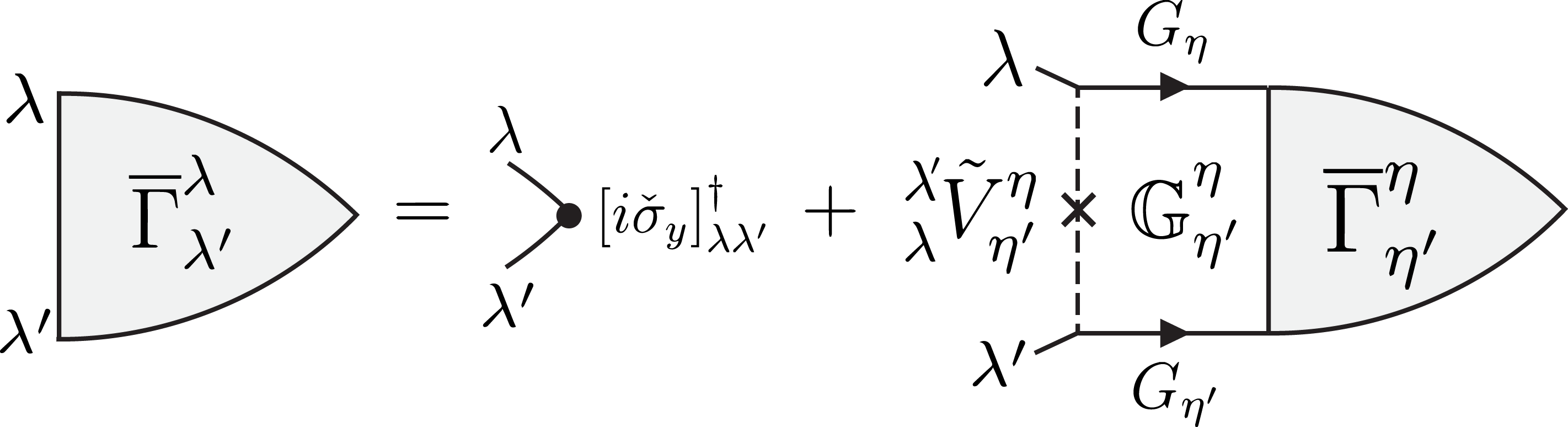}
    \caption{Diagrammatic presentation of the integral equation \eqref{eq:Cooperon} satisfied by the Cooperon vertex, $\overline{ \Gamma }$. 
    The disorder scattering vertex, $\tilde{V}$, is defined in Eq.~\eqref{eq:V_tilde}, and $\mathbb{G}$ stands for the product of the two Green functions, see Eq.~\eqref{eq:GF_pair}.}
    \label{fig:GammaEq}
\end{figure}

The Cooperon vertex introduced in Eq.~\eqref{eq:Bubble_w_Cooperon} satisfies the integral equation (see Fig.~\ref{fig:GammaEq})
\begin{align}\label{eq:Cooperon}
	\overline{ \Gamma }^{ \lambda }_{ \lambda' } & \left( \mathbf{k}, \mathbf{q}, \mathbf{B}, \varepsilon_n \right)
	   =
	\left[ i \check{\sigma}_y \right]^\dagger_{ \lambda\lambda' }
  +  
	\sum_{ \mathbf{p}' \eta \eta' } 
	{}^{\lambda}_{\lambda'}\!\tilde{V}^{\eta}_{\eta'} \left( \mathbf{k},\mathbf{k}' \right) 
 \notag \\
& \times	 \mathbb{G}^{ \eta }_{ \eta' } \left(  \mathbf{k}',  \mathbf{q},  \mathbf{B}, \varepsilon_n \right)
	\overline{ \Gamma }^{ \eta }_{ \eta' } \left(  \mathbf{k}',  \mathbf{q},  \mathbf{B}, \varepsilon_n \right) \,,
\end{align}
where $\tilde{V}$ is the scattering vertex in the chiral basis. 
All the important electron momenta are close to $k_F$.
For such momenta we have
\begin{align}\label{eq:V_tilde}
    {}^{\lambda}_{\lambda'}\!\tilde{V}^{\eta}_{\eta'}  \left( \mathbf{k},\mathbf{k}' \right) 
    \approx & \frac{V^2}{4} \left[  1 + \lambda \eta e^{ - i \left( \varphi - \varphi' \right) } 
    \right]
    \notag \\
    & \times 
    \left[ 1 + \lambda' \eta' e^{ - i \left( \varphi - \varphi' \right) } \right]\, .
\end{align}
This equation holds under the same conditions as Eq.~\eqref{eq:check_sigma}, see Appendix~\ref{app:disorder}.

As discussed in Appendix~\ref{app:inter-band_pairing} the inter-band scattering processes captured by Eq.~\eqref{eq:Cooperon} by the terms with $\eta \neq \eta'$ give a negligible contribution under the condition \(\max\{1/\tau,T_c \} \ll \Delta_{\mathrm{SO}}\). 
This restriction is consistent with the Green function~\eqref{eq:G_lambda} being diagonal in the band index.
The above condition, in fact ensures the two spin split bands can be treated as a two-band superconductor.
In this approximation the inter-band pairing has a negligible effect.

Considering only the intra-band Cooper pairs we follow the standard route by integrating over fast electron momentum introducing the integral, 
\begin{align}\label{eq:GG_bar}
	\overline{ \mathbb{G} }_{ \lambda }^{ \lambda } \left( \varphi, \mathbf{q}, \mathbf{B}, \varepsilon_n \right)
    	= &
    	\nu_\lambda \int_{-\infty}^{\infty} d \xi_{ \mathbf{k} }^\lambda   \mathbb{G}_{ \lambda }^{ \lambda } \left( \mathbf{k}, \mathbf{q}, \mathbf{B}, \varepsilon_n \right)
        \notag \\
         = &  \frac{2 \pi \nu_\lambda}{2 |\varepsilon_n| + 2 i \mathbf{v}_F \cdot \mathbf{Q}^\lambda + \frac{\mathrm{sgn}(\varepsilon_n)}{\tau}}.
\end{align}

The integration in Eq.~\eqref{eq:GG_bar} has an effect of setting all the electron momenta to the Fermi momentum.
In result, keeping only intra-band contributions, Eq.~\eqref{eq:Cooperon} is transformed to 
\begin{align}\label{eq:Cooperon_a}
	\overline{ \Gamma }^{ \lambda }_{ \lambda} & \left( \varphi, \mathbf{q}, \mathbf{B}, \varepsilon_n \right)
	   =
	\left[ i \check{\sigma}_y \left( \varphi \right)\right]^\dagger_{ \lambda\lambda }
  +  
	\int \frac{d \varphi'}{2 \pi}
        \sum_{ \eta } 
	{}^{\lambda}_{\lambda}\!\tilde{V}^{\eta}_{\eta} \left( \varphi- \varphi' \right) 
 \notag \\
& \times	\overline{ \mathbb{G} }^{ \eta }_{ \eta } \left( \varphi',  \mathbf{q},  \mathbf{B}, \varepsilon_n \right)
	\overline{ \Gamma }^{ \eta }_{ \eta } \left(  \varphi',  \mathbf{q},  \mathbf{B}, \varepsilon_n \right) \, .
\end{align}

\begin{figure}
    \centering
    \includegraphics[width = 0.48\textwidth]{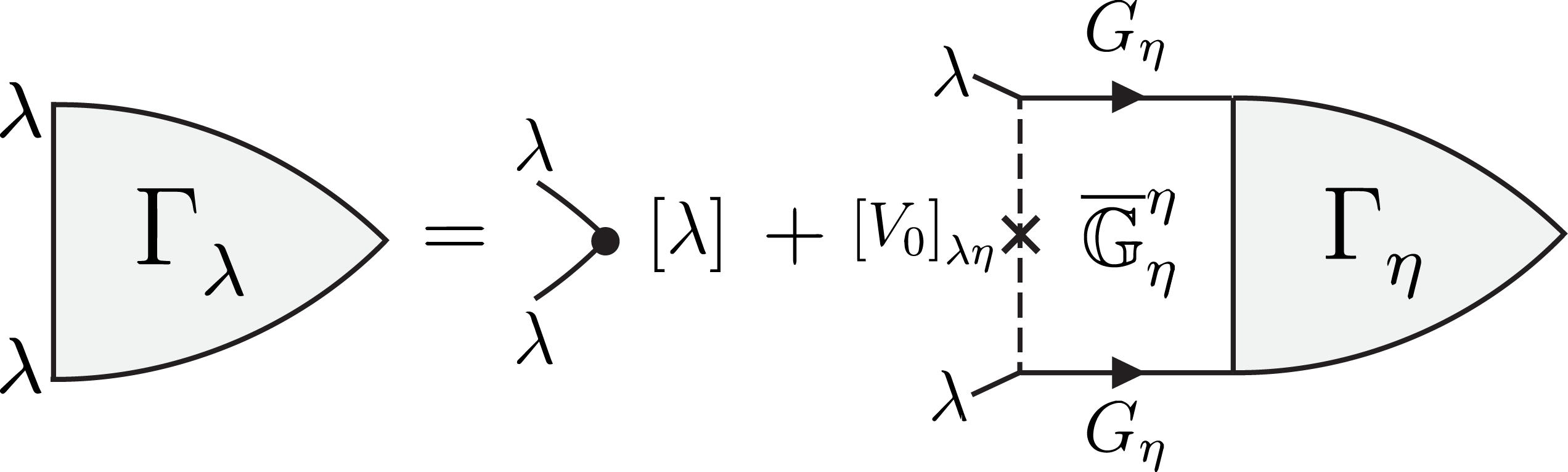}
    \caption{Diagrammatic representation of the integral equation \eqref{eq:Cooperon_b} satisfied by the modified Cooperon vertex, $\Gamma$.
    The disorder scattering amplitude $V_0$ is defined by Eq.~\eqref{eq:V0_a}.
    The integrated product of the two Green functions, $\bar{\mathbb{G}}$ is given by Eq.~\eqref{eq:GG_bar}.}
    \label{fig:Cooperon_b}
\end{figure}

It is convenient to introduce the modified vertex function
\begin{align}\label{eq:Gamma_lambda}
    \Gamma_\lambda(\varphi) = i e^{i \varphi} \overline{ \Gamma }^{ \lambda }_{ \lambda}  (\varphi)
\end{align}
and the modified disorder scattering vertex,
\begin{align}\label{eq:V0}
	{}^{ \lambda }_{ \lambda }{\tilde{V}}^{ \eta }_{ \eta } \left( \varphi - \varphi' \right)
	=
	{ \left[ V_0 \right] }_{\lambda \eta } \left( \varphi - \varphi' \right)\, .
	\left[ e^{ - i (\varphi - \varphi')} \right] \, .
\end{align}
Multiplying Eq.~\eqref{eq:Cooperon_a} by \( i e^{ i \varphi} \), and using the definitions \eqref{eq:Gamma_lambda} and \eqref{eq:V0} we write it in the form,
\begin{align}\label{eq:Cooperon_b}
	\Gamma_{ \lambda } \left( \varphi \right)
	 = &
	\lambda
	+
	\sum_{ \eta }
	\int_{ 0 }^{ 2 \pi } \frac{d \varphi'}{2 \pi}
	{ \left[ V_0 \right] }_{\lambda\eta} \left( \varphi - \varphi' \right)
 \notag \\
& \times 	\overline{\mathbb{G}}^{ \eta }_{ \eta } \left( \varphi' \right)
	\Gamma_{ \eta } \left( \varphi' \right)  \, , 
\end{align}
illustrated graphically in Fig.~\ref{fig:Cooperon_b}.
The disorder scattering vertex $V_0$ in Eq.~\eqref{eq:Cooperon_b} is fixed by Eqs.~\eqref{eq:V_tilde} and \eqref{eq:V0}, 
\begin{align}\label{eq:V0_a}
    { \left[ V_0 \right] }_{\lambda\eta} = \frac{V^2}{2} \left[\cos( \varphi - \varphi' ) + \lambda \eta \right]\, ,
\end{align}
see Appendix~\ref{app:disorder_mod}.

It is clear from Eq.~\eqref{eq:Cooperon_b} and the form of the disorder scattering amplitude~\eqref{eq:V0_a} that the solution to the Eq.~\eqref{eq:Cooperon_b} takes the form 
\begin{align}\label{eq:Gamma_harmonics}
	\Gamma_{ \lambda } \left( \varphi \right)
    	& = C_{ \lambda } ^{ \left( 0 \right) } 
    	+ C_{ \lambda } ^{ \left( c \right) } \cos \varphi 
    	+ C_{ \lambda } ^{ \left( s \right) } \sin  \varphi \, .
\end{align}
The specific form of Eq.~\eqref{eq:Gamma_harmonics} turns equation \eqref{eq:Cooperon_b} into six linear algebraic equations for the six unknown coefficients $C_{ \pm } ^{ \left( 0 \right) }$, 
$C_{ \pm } ^{ \left( c \right) } $, and $C_{ \pm } ^{ \left( s \right) } $.
These equations can be summarized as follows.
Introduce the column vector 
\begin{align}\label{eq:C_vec}
    \mathbf{C}= \left[C_{ + }^{ \left( 0 \right) }, C_{ + }^{ \left( c \right) },C_{ + }^{ \left( s \right) },C_{ - }^{ \left( 0 \right) } , C_{ - }^{ \left( c \right) } ,C_{ - }^{ \left( c \right) }\right]^{\mathrm{t}}\, ,
\end{align}
where the superscript $\mathrm{t}$ stands for the transposition.
The linear equation satisfied by $\mathbf{C}$ takes the form,
\begin{align}\label{eq:M}
	\left( \mathbb{M} - \mathbb{1}_6 \right) \mathbf{C} =
	\left[- 1,  0,  0,  1,  0,   0 \right]^{\mathrm{t}}\, ,
\end{align}
where $\mathbb{1}_6 $ is the \(6\times6\) unit matrix, and $\mathbb{M}$ can be written as the \(6\cross6\) matrix 
\begin{align}\label{eq:M_matrix}
    \mathbb{M}
	= 
    	\mqty[
    	A_+^{00} & A_+^{10} & A_+^{01} & - A_-^{00} & - A_-^{10} & -A_-^{01}
    	\\
    	A_+^{10} & A_+^{20} & A_+^{11} & A_-^{10} & A_-^{20} & A_-^{11}
    	\\
    	A_+^{01} & A_+^{11}  & A_+^{02}  & A_-^{01} & A_-^{11} & A_-^{02}
    	\\
    	- A_+^{00} & - A_+^{10} & - A_+^{01} & A_-^{00}  & A_-^{10} & A_-^{01}
    	\\
    	A_+^{10} & A_+^{20} & A_+^{11} & A_-^{10} & A_-^{20}  & A_-^{11}
    	\\
    	A_+^{01} & A_+^{11} & A_+^{02} & A_-^{01} & A_-^{11} & A_-^{02} 
    	] \,,
\end{align}
where 
\begin{align}\label{eq:A_def}
	A_\lambda^{ m n } 
	= \frac{V^2 }{ 2} \int_0^{2 \pi} \frac{ d \varphi }{2 \pi} \overline{ \mathbb{G} }^{ \lambda }_{ \lambda } \left( \varphi \right) \cos^m \varphi  \sin^n \varphi  \, .
\end{align}
A more detailed derivation of Eqs.~\eqref{eq:C_vec}-\ref{eq:A_def} is given in Appendix~\ref{app:Cooperon_matrix}.

\begin{figure}
    \centering
    \includegraphics[width = 0.48\textwidth]{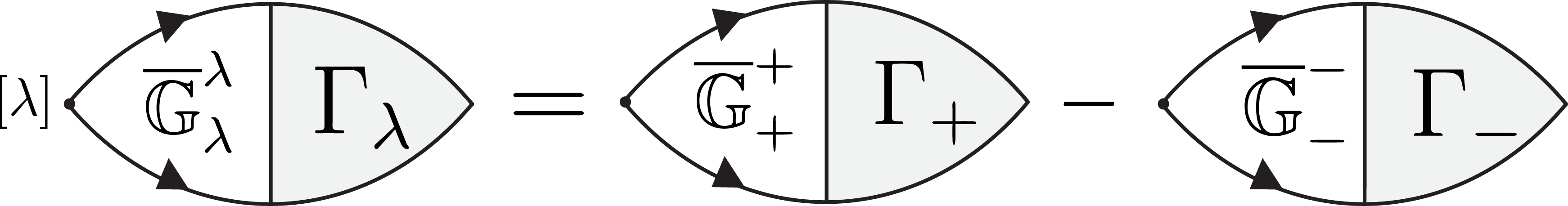}
    \caption{Diagrammatic presentation of the expression \eqref{eq:Pi_end} for the polarization operator \( \Pi \left( \mathbf{q} \right) \).
    The integrated product of the two Green functions $\bar{\mathbb{G}}$ is given by Eq.~\eqref{eq:GG_bar}. The Cooperon vertex, $\Gamma$, satisfies the integral Eq.~\eqref{eq:Cooperon_b} illustrated in Fig.~\ref{fig:Cooperon_b}.}
    \label{fig:Pi_end}
\end{figure}

To complete the calculation, we rewrite the expression for the polarization operator~\eqref{eq:Bubble_w_Cooperon} within the approximations made above in the form (see Fig.~\ref{fig:Pi_end})
\begin{align}\label{eq:Pi_end}
\Pi \left( \mathbf{q} \right) =    
    2 T \sum_{ \varepsilon_n } \int_{0}^{2 \pi} \frac{d \varphi}{2 \pi}
		\left(
				\overline{ \mathbb{G} }^{ + }_{ + } 
				\Gamma_{ + } 
			-
				\overline{ \mathbb{G} }^{ - }_{ - } 
				\Gamma_{ - }
            \right)\, .
\end{align}

We have checked that at $q=0$ and $B=0$, Eq.~\eqref{eq:Pi_end} reduces to the expression, \( \Pi \left( \mathbf{q} \right)  = 4 \pi \nu_0 /|\varepsilon_n| \) which does not include the disorder in accordance with the Anderson theorem. 

\begin{figure}
    \centering
    \includegraphics[width = 0.48\textwidth]{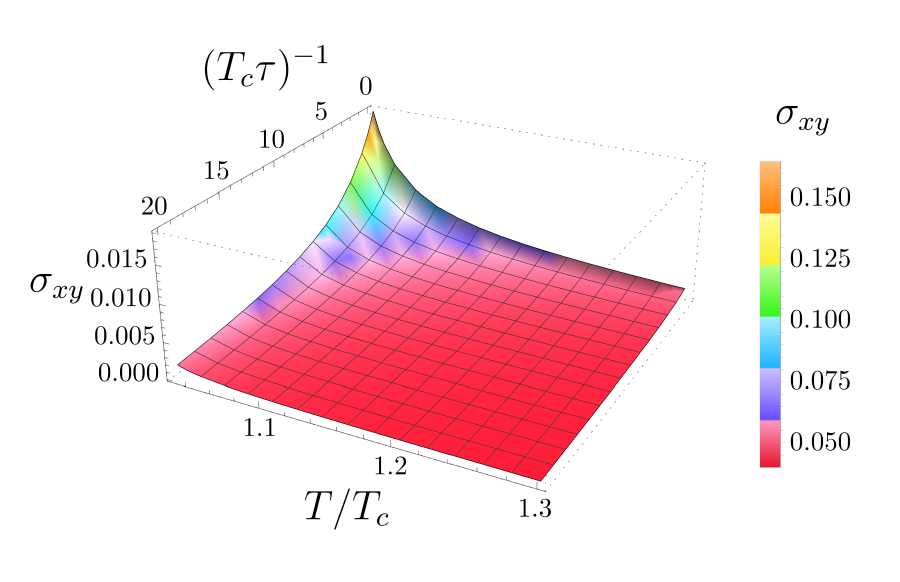}
    \caption{(color online) Planar Hall conductance \( \sigma_{xy} \) as a function of disorder strength \( 1/\tau \) and temperature $T$ both shown in units of \(T_c \). 
    The conductance is given in units of \( B_x B_y / 16 e^2 T_c^2 \). 
    The conductance is a monotonically decreasing function of both variables with reaching the maximum in the clean limit.}
    \label{fig:sigmaXY}
\end{figure}

The outline of the remaining calculation is as follows.
We solve Eq.~\eqref{eq:M} which gives the Cooperon vertex via the expansion, Eq.~\eqref{eq:Gamma_harmonics} (detailed in Appendix~\ref{app:Cooperon_solution}).
This solution, in turn, allows us to evaluate the polarization operator using Eq.~\eqref{eq:Pi_end}.
The knowledge of $\Pi(\mathbf{q})$ translates directly into the Cooper pair dispersion, Eq.~\eqref{eq:pair_propagator}. 
This part of the calculation is detailed in Appendix~\ref{app:polarization_bubble}.
The results of the calculation are represented in Fig.~\ref{fig:sigmaXY}.

\section{Discussion and outlook}
\label{sec:conclusion}

We have investigated how superconducting fluctuations induce a two fold magneto resistance anisotropy in the form of the PHE.
While the PHE vanishes in the absence of interactions, pairing in the Cooper channel makes PHE finite.
This makes the paraconductivity due to the superconducting fluctuations the primary source of the two-fold magneto-resistance anisotropy in 2D systems with broken basal plane mirror symmetry, $\sigma_h$.
Based on symmetry, we have constructed the phenomenology of the PHE.
The key observation is that in the planar configuration, the weak field PHE requires breaking of the $\sigma_h$ symmetry.

We note that in non-planar configurations, $\sigma_h$ symmetry breaking is not necessary for transport anisotropies.
Non-linear anisotropic paraconductivity has been reported to result in a strongly non-reciprocal current-voltage characteristic of gated MoS$_2$ \cite{Wakatsuki2017}.
Even though this 2D system possesses $\sigma_h$ symmetry, the perpendicular magnetic field couples to a thermal and non-equilibrium Cooper pairs causing the transport non-reciprocity in the normal state.

Our findings bear implications on the PHE recently observed in the topological surface states \cite{Taskin2017}.
Our derivation applies directly to the metallic regime, where the Fermi energy is far from the Dirac point.
At charge neutrality, however, the PHE in the non-interacting regime is appreciable \cite{Trushin2009,Wang2020}.
Yet, even close to Dirac point the superconducting fluctuations may modify the PHE.
In particular, triplet correlations have the potential to play an important role, as the spin-orbit coupling is the dominant energy scale close to charge neutrality.

We stress that the PHE describes the anisotrpic, yet reciprocal transport.
It is, therefore, distinct from the superconducting diode effect (SDE) which recently attracts considerable attention \cite{Ando2020,Yuan2022,Kochan2023}.
The PHE and SDE are expressed via the terms in the Ginzburg-Landau functional that are even and odd in momentum, respectively. 
More specifically, the PHE arises from the field-induced modification of the pair dispersion relation in the second order of pair momentum.
At the same time, in a planar geometry, the first-order terms, \(\mathcal{L}_1\), known as Lifshitz invariants, appear together with the second-order terms, \(\mathcal{L}_2\).
For this reason consistency requires the consideration of both the first and second order terms.
The Lifshitz invariants play an essential role in the SDE, and no role in PHE.

Once we have clarified the symmetry requirements for the PHE, we focused on the  2D Rashba superconductor in the planar configuration (the so-called Rashba-Zeeman superconductor).
By definition Rashba spin-orbit coupling breaks $\sigma_h$ symmetry.
We have computed the PHE within this model for a wide range of parameters.
The calculation is performed for spin-orbit interaction exceeding the critical temperature in both clean and dirty limits. 
Certainly, the PHE is finite in all regimes, and we have considered the specific limit for concreteness (Fig.~\ref{fig:sigmaXY}).
We argue that other types of spin-orbit interactions breaking the $\sigma_h$ symmetry such as the Dresselhaus spin-orbit interaction yield very similar results to what we have found in the case of Rashba superconductor.

We make, in fact, a more general observation regarding the $\sigma_h$ symmetry.
Any 2D system in the planar configuration (Fig.~\ref{fig:plane}) is totally isotropic with respect to the magnetic field orientation unless $\sigma_h$ symmetry is broken.
Indeed, two identical systems differing by the orientation of the in-plane field are unitarily equivalent.
This implies, in particular, that if $\sigma_h$ is not broken the critical field remains totally isotropic despite the hexagonal symmetry of the underlying lattice in few-layer NbSe$_2$ \cite{Hamill2021}.
In this case the six-fold symmetry is in fact not expected.
In contrast, the six-fold critical field variation in mono-layers of NbSe$_2$ \cite{Cho2022} is a strong indication of the broken $\sigma_h$ symmetry in the form of a Rashba spin-orbit interaction.

Although we mainly focused on the part of the pair dispersion which is second order in momentum, we also reproduced the results for the first order, i.e. the Lifshitz invariants.
In addition, in some limited range of parameters we have found other contributions to Lifshitz invariants that have been overlooked so far, see the discussion that follows Eqs.~\eqref{eq:deltaL1} and \eqref{eq:deltaL1_triplet}.
These contributions stem from 1) the field induced modification of the interaction matrix elements; 2) the field induced coupling of singlet and $p$-wave triplet-paring channels. 
This not only serves as a benchmark for our findings but also holds relevance for studies of SDE.

Finally, we speculate that the field induced anisotropy of the Cooper pair dispersion may have a clear experimental signature in the vortex state.
One might expect an elliptical shape of the vortex core with the ellipticity controlled by the in-plane component of the magnetic field, and proportional to $\mathcal{L}_2$.
Very recently, the field induced anisotropy of the pinning force has been reported \cite{Fuchs2022}.
It has been analyzed in terms of the Lifshitz invariant in a Rashba-Zeeman superconductor.
It is plausible that the terms quadratic in momentum investigated in this study directly influence this anisotropy. 
This offers a distinct yet related avenue of research.

\section*{Acknowledgements}
We thank  Y. Dagan, M. Dzero, A. Levchenko, and D. M\"ockli for useful discussions. 
L.A. and M.K. acknowledge the financial support from the Israel Science Foundation, Grant No. 2665/20. 
\appendix

\section{Pairing interaction Hamiltonian in the basis, \texorpdfstring{Eq.~\eqref{eq:psiB}}{Eq.9}}
\label{app:inter-band_scattering}

In this appendix we transform the singlet-pairing interaction Hamiltonian~\eqref{eq:Hps}, to the basis that diagonalizes the free part of the Hamiltonian, Eq.~\eqref{eq:psiB}.
We then show that to the zero-order expansion in \( q/k_F, B/\Delta_{\mathrm{SO}} \ll 0 \), the interaction vertex is given by Eq.~\eqref{eq:check_sigma}.
This procedure amounts to a transformation to the chiral basis, Eq.~\eqref{eq:chiral_basis}.
For completeness the transformation is preformed in two steps: first transform to the basis of Eq.~\eqref{eq:psiB}, and second reduce it to the transformation to the basis of Eq.~\eqref{eq:chiral_basis}.

The inverse transformation Eq.~\eqref{eq:psiB_inv} implies
\begin{subequations}\label{cd_spin_operators}
	\begin{align}
		\hat{c}^\dag_\uparrow & \left( \vb{k} + \frac{\vb{q}}{2} \right) 
		 \!=\! \frac{ 1 }{ \sqrt{2} } \left[ \hat{ \psi }_{\mathbf{B}}^{+\dag}\left( \vb{k} + \frac{\vb{q}}{2} \right)\! +\! \hat{ \psi }_{\mathbf{B}}^{-\dag}\left( \vb{k} + \frac{\vb{q}}{2} \right) \right] 
		\\
		\hat{c}^\dag_\downarrow & \left( \vb{k} + \frac{\vb{q}}{2} \right) 
		 = \frac{ 1 }{ \sqrt{2} } \left[ \hat{ \psi }_{\mathbf{B}}^{+\dag}\left( \vb{k} + \frac{\vb{q}}{2} \right) - \hat{ \psi }_{\mathbf{B}}^{-\dag}\left( \vb{k} + \frac{\vb{q}}{2} \right) \right] 
\notag \\		
 & \times \left[
		\frac{ \left( k_y + \frac{  q_y }{2} + \frac{ B_x }{ \alpha } \right) - i \left( k_x + \frac{ q_x }{ 2 } - \frac{ B_y }{ \alpha } \right) }
		{ \sqrt{ \left( k_x + \frac{ q_x }{ 2 } - \frac{ B_y }{ \alpha } \right)^2 + \left( k_y + \frac{  q_y }{2} + \frac{ B_x }{ \alpha } \right)^2 } } 
		\right]
	\end{align}
\end{subequations}
and similarly
\begin{subequations}\label{c_spin_operators}
	\begin{align}
		\hat{c}_\uparrow \! &\! \left( \vb{k}' + \frac{\vb{q}}{2} \right) 
		 = \frac{ 1 }{ \sqrt{2} } \left[ \hat{ \psi }^+_\mathbf{B}\left( \vb{k}' \!+\! \frac{\vb{q}}{2} \right) + \hat{ \psi }^-_\mathbf{B}\left( \vb{k}' \!+\! \frac{\vb{q}}{2} \right) \right] 
		\\
		\hat{c}_\downarrow & \left( \vb{k}' + \frac{\vb{q}}{2} \right) 
		 = \frac{ 1 }{ \sqrt{2} } \left[ \hat{ \psi }^+_\mathbf{B}\left( \vb{k}' + \frac{\vb{q}}{2} \right) - \hat{ \psi }^-_\mathbf{B}\left( \vb{k}' + \frac{\vb{q}}{2} \right) \right]
   \notag \\
	& \times	\left[ 
		\frac{ \left( k_y' + \frac{  q_y }{2} + \frac{ B_x }{ \alpha } \right) + i \left( k_x' + \frac{ q_x }{ 2 } - \frac{ B_y }{ \alpha } \right) }
		{ \sqrt{ \left( k_x' + \frac{ q_x }{ 2 } - \frac{ B_y }{ \alpha } \right)^2 + \left( k_y' + \frac{  q_y }{2} + \frac{ B_x }{ \alpha } \right)^2 } } 
		\right].
	\end{align}
\end{subequations}
We therefore have, based on Eq.~\eqref{cd_spin_operators},
\begin{align}\label{eq:tr1}
		& \hat{c}^\dag_\uparrow \left( \vb{k} + \frac{\vb{q}}{2} \right) 
		\left[ i \sigma_y \right]_{\uparrow \downarrow} 
		\hat{c}^\dag_\downarrow \left( - \vb{k} + \frac{\vb{q}}{2} \right)
\notag \\		
  & = \frac{1}{2}
		\left[ \hat{ \psi }^{+ \dag}_\mathbf{B}\left( \vb{k} + \frac{\vb{q}}{2} \right) 
  + \hat{ \psi }^{- \dag}_\mathbf{B}\left( \vb{k} + \frac{\vb{q}}{2} \right) \right]
  \notag \\
& \times 		\left[ \hat{ \psi }^{+ \dag}_\mathbf{B}\left( - \vb{k} + \frac{\vb{q}}{2} \right) - \hat{ \psi }^{- \dag}_\mathbf{B}\left( - \vb{k} + \frac{\vb{q}}{2} \right) \right] 
		\nonumber
		\\
		& 
	\times	\left[
		\frac{ \left( - k_y + \frac{  q_y }{2} + \frac{ B_x }{ \alpha } \right) - i \left( - k_x + \frac{ q_x }{ 2 } - \frac{ B_y }{ \alpha } \right) }
		{ \sqrt{ \left( - k_x + \frac{ q_x }{ 2 } - \frac{ B_y }{ \alpha } \right)^2 + \left( - k_y + \frac{  q_y }{2} + \frac{ B_x }{ \alpha } \right)^2 } } 
		\right]
\end{align}
and similarly, 
\begin{align}\label{eq:tr2}
& \hat{c}^\dag_\downarrow \left( \vb{k} + \frac{\vb{q}}{2} \right)
		\left[ i \sigma_y \right]_{\downarrow \uparrow } 
		\hat{c}^\dag_\uparrow \left( - \vb{k} + \frac{\vb{q}}{2} \right) 
\notag \\		
&  =- \frac{1}{2} 
		\left[ \hat{ \psi }^{+ \dag}_\mathbf{B}\left( \vb{k} + \frac{\vb{q}}{2} \right) 
  - \hat{ \psi }^{- \dag}_\mathbf{B}\left( \vb{k} + \frac{\vb{q}}{2} \right) \right]
  \notag \\
		& \times \left[ \hat{ \psi }^{+ \dag}_\mathbf{B}\left( - \vb{k} + \frac{\vb{q}}{2} \right) + \hat{ \psi }^{- \dag}_\mathbf{B}\left( - \vb{k} + \frac{\vb{q}}{2} \right) \right]
		\nonumber\\
		& \times \left[
		\frac{ \left( k_y + \frac{  q_y }{2} + \frac{ B_x }{ \alpha } \right) - i \left( k_x + \frac{ q_x }{ 2 } - \frac{ B_y }{ \alpha } \right) }
		{ \sqrt{ \left( k_x + \frac{ q_x }{ 2 } - \frac{ B_y }{ \alpha } \right)^2 + \left( k_y + \frac{  q_y }{2} + \frac{ B_x }{ \alpha } \right)^2 } } 
		\right].
\end{align}
The sum of the two contributions~\eqref{eq:tr1} and \eqref{eq:tr2} gives rise to the vertex in the basis, Eq.~\eqref{eq:psiB},
\begin{align}\label{SC_Pauli_chiral}
	& \left[ i \check{\sigma}_y \left( \vb{k}, \vb{q}, \vb{B} \right) \right]_{ \lambda \lambda' }  
		=
  \notag \\
		& - \frac{1}{2} \left[
		\lambda
		\frac{ \left( k_y + \frac{  q_y }{2} + \frac{ B_x }{ \alpha } \right) - i \left( k_x + \frac{ q_x }{ 2 } - \frac{ B_y }{ \alpha } \right) }
		{ \sqrt{ \left( k_x + \frac{ q_x }{ 2 } - \frac{ B_y }{ \alpha } \right)^2 + \left( k_y + \frac{  q_y }{2} + \frac{ B_x }{ \alpha } \right)^2 } } 
  \right.
		\notag \\
		-
		&  \lambda'		\left.	
		\frac{ \left( - k_y + \frac{  q_y }{2} + \frac{ B_x }{ \alpha } \right) - i \left( - k_x + \frac{ q_x }{ 2 } - \frac{ B_y }{ \alpha } \right) }
		{ \sqrt{ \left(  k_x - \frac{ q_x }{ 2 } + \frac{ B_y }{ \alpha } \right)^2 + \left(  k_y - \frac{  q_y }{2} - \frac{ B_x }{ \alpha } \right)^2 } } 
	\right].
\end{align}
Repeating the same steps for the annihilation operators, Eq.~\eqref{c_spin_operators} we obtain the result, 
\begin{align}\label{SC_Pauli_chiral_hc}
	& \left[ i \check{\sigma}_y \left( \vb{k}', \vb{q}, \vb{B} \right) \right]_{ \lambda \lambda' }^\dag  
		=
  \notag \\
		& - \frac{1}{2} \left[
		\lambda
		\frac{ \left( k'_y + \frac{  q_y }{2} + \frac{ B_x }{ \alpha } \right) + i \left( k'_x + \frac{ q_x }{ 2 } - \frac{ B_y }{ \alpha } \right) }
		{ \sqrt{ \left( k'_x + \frac{ q_x }{ 2 } - \frac{ B_y }{ \alpha } \right)^2 + \left( k'_y + \frac{  q_y }{2} + \frac{ B_x }{ \alpha } \right)^2 } } 
  \right.
		\notag \\
		-
		&  \lambda'		\left.	
		\frac{ \left( - k'_y + \frac{  q_y }{2} + \frac{ B_x }{ \alpha } \right) + i \left( - k'_x + \frac{ q_x }{ 2 } - \frac{ B_y }{ \alpha } \right) }
		{ \sqrt{ \left(  k'_x - \frac{ q_x }{ 2 } + \frac{ B_y }{ \alpha } \right)^2 + \left(  k'_y - \frac{  q_y }{2} - \frac{ B_x }{ \alpha } \right)^2 } } 
	\right].
\end{align}
\subsection{Approximate expression for the interaction amplitude}\label{app:inter-band_zero_order}
As explained in Sec.~\ref{sec:chiral}, in the regime considered in this work it is sufficient to keep the interaction vertices, Eqs.~\eqref{SC_Pauli_chiral} and \eqref{SC_Pauli_chiral_hc}, to zero order in $q/k_F$ and $B/\Delta_{\mathrm{SO}}$.
Writing the momentum $\mathbf{k} = k \hat{x} \cos \varphi +k \hat{y} \sin \varphi $ as in the main text, we reduce the interaction vertices to the following expressions: 
\begin{subequations}\label{sigmay_0}
	\begin{align}
		\left[i \check{\sigma}_y \left(  \varphi \right) \right]
		& = i e^{ i \varphi } \sigma_z
		\label{sigmay_0nd}
		\\
		\left[ i \check{\sigma}_y \left( \varphi' \right) \right]^\dag 
		& = -i e^{ - i \varphi' } \sigma_z
		\label{sigmay_0d}
		\, .
	\end{align}
\end{subequations}
Crucially, the interaction amplitudes in the chiral representation are purely intra-band.
%%%
\section{Disorder scattering vertex}
\label{app:disorder}
Here we transform the disorder Hamiltonian, and the disorder scattering vertex to the basis of Eq.~\eqref{eq:psiB}.
As in the case of the pairing interaction, Eq.~\eqref{sigmay_0}, we keep zero order in both $q/k_F$ and $B/\Delta_{\mathrm{SO}}$ in the scattering vertices, thus greatly simplifying them. 

To write the spin conserving Hamiltonian~\eqref{eq:Hdis} in the basis Eq.~\eqref{eq:psiB}, it is enough to use the following transformed bilinear combinations,
\begin{align}\label{scatter_up}
	\hat{c}^\dag_\uparrow & \left( \vb{k}' + \vb{q}/2 \right) \hat{c}_\uparrow \left( \vb{k} + \vb{q}/2 \right) 
 \notag \\
    	 & = \frac{1}{2} \sum_{ \eta, \eta' = +, - } 
        \hat{ \psi }_{\mathbf{B}}^{\eta \dag}\left( \vb{k}' + \frac{\vb{q}}{2} \right)
        \hat{ \psi }_{\mathbf{B}}^{\eta'}\left( \vb{k} + \frac{\vb{q}}{2} \right)
        \, ,
\end{align} 
and similarly,
\begin{align}\label{scatter_down}
	\hat{c}^\dagger_\downarrow & \left( \vb{k}' + \vb{q}/2 \right) 
    \hat{c}_\downarrow \left( \vb{k} + \vb{q}/2 \right)
        \notag \\
        = & 
        \frac{1}{2} \sum_{ \eta, \eta' = +, - } \left( \eta \eta' \right)
        \hat{ \psi }_{\mathbf{B}}^{\eta \dag}\left( \vb{k}' + \frac{\vb{q}}{2} \right)
        \hat{ \psi }_{\mathbf{B}}^{\eta'}\left( \vb{k} + \frac{\vb{q}}{2} \right)
        \notag \\
        & \times
			\frac{ \left( k_y' + \frac{ q_y }{2} + \frac{ B_x }{ \alpha } \right) - i \left( k_x' + \frac{ q_x }{ 2 } - \frac{ B_y }{ \alpha } \right) }{ \sqrt{ \left( k_x' + \frac{ q_x }{ 2 } - \frac{ B_y }{ \alpha } \right)^2 + \left( k_y' + \frac{  q_y }{2} + \frac{ B_x }{ \alpha } \right)^2 } } 
        \notag \\		
  	&	\times  
		      \frac{ \left( k_y + \frac{ q_y }{2} + \frac{ B_x }{ \alpha } \right) + i \left( k_x + \frac{ q_x }{ 2 } - \frac{ B_y }{ \alpha } \right) } { \sqrt{ \left( k_x + \frac{ q_x }{ 2 } - \frac{ B_y }{ \alpha } \right)^2 + \left( k_y + \frac{  q_y }{2} + \frac{ B_x }{ \alpha } \right)^2 }} 	\,.
\end{align}
The Hamiltonian~\eqref{eq:Hdis} takes the form
\begin{align}\label{eq:Hdis1}
	\mathcal{H}_{ \text{dis} } 
        = &
	   \sum_{\mathbf{R}_j} \sum_{ \vb{k}', \eta, \eta' } V_{ \eta \eta' } \left( \vb{k}, \vb{k}'; \vb{q}, \vb{B} \right) e^{ i (\mathbf{k}' - \mathbf{k} )\cdot \mathbf{R}_j}
\notag \\
     & \times 
        \hat{ \psi }_{\mathbf{B}}^{\eta \dag}\left( \vb{k}' + \frac{\vb{q}}{2} \right)
        \hat{ \psi }_{\mathbf{B}}^{\eta'}\left( \vb{k} + \frac{\vb{q}}{2} \right)
        \, ,
\end{align}
where 
\begin{align}\label{chiral_disorder}
	V_{ \eta \eta' } & \left( \vb{k}, \vb{k}'; \vb{q}, \vb{B} \right) = \frac{V}{2}
\notag \\
 & + \frac{V}{2} \eta \eta' 
	\frac{ \left( k_y' + \frac{  q_y }{2} + \frac{ B_x }{ \alpha } \right) - i \left( k_x' + \frac{ q_x }{ 2 } - \frac{ B_y }{ \alpha } \right) }
	{ \sqrt{ \left( k_x' + \frac{ q_x }{ 2 } - \frac{ B_y }{ \alpha } \right)^2 + \left( k_y' + \frac{  q_y }{2} + \frac{ B_x }{ \alpha } \right)^2 } } 
 \notag \\
& \times 
	\frac{ \left( k_y + \frac{  q_y }{2} + \frac{ B_x }{ \alpha } \right) + i \left( k_x + \frac{ q_x }{ 2 } - \frac{ B_y }{ \alpha } \right) }
	{ \sqrt{ \left( k_x + \frac{ q_x }{ 2 } - \frac{ B_y }{ \alpha } \right)^2 + \left( k_y + \frac{  q_y }{2} + \frac{ B_x }{ \alpha } \right)^2 } } .
\end{align}

\begin{figure}
    \centering
    \includegraphics[scale = 0.40]{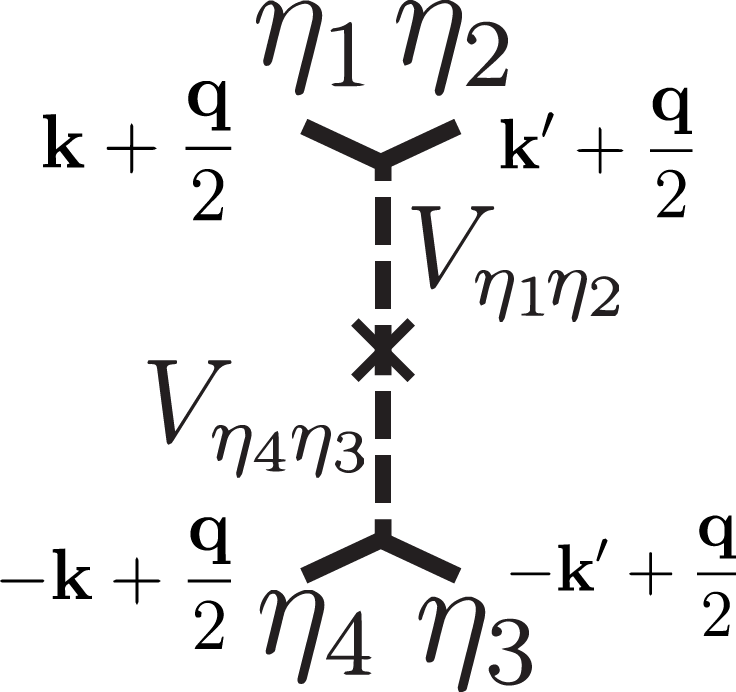}
    \caption{Diagrammatic presentation of the disorder scattering vertex~\eqref{V2}.
    The matrix elements $V_{\eta_i \eta_j}$ are given by Eq.~\eqref{chiral_disorder}.}
    \label{fig:V2}
\end{figure}

The disorder scattering amplitude is schematically shown in Fig.~\ref{fig:V2}.
Based on Eqs.~ \eqref{eq:Hdis1} and \eqref{chiral_disorder} it takes the form 
\begin{align}\label{V2}
	{}^{\eta_1}_{\eta_4}\! {\tilde{V}}^{ \eta_2 }_{ \eta_3 }  \left( \vb{k}, \vb{k}'; \vb{q, \vb{B}} \right)
	= & V_{ \eta_1 \eta_2 } \left( \vb{k}, \vb{k}'; \vb{q}, \vb{B} \right) 
 \notag \\
& \times  V_{ \eta_4 \eta_3 } \left( - \vb{k}, - \vb{k}'; \vb{q}, \vb{B} \right).
\end{align} 
To zero order in $q/k_F$ and $B/\Delta_{\mathrm{SO}}$, Eq.~\eqref{chiral_disorder} is simplified to 
\begin{align}\label{eq:chiral_disorder_s}
	V_{ \eta \eta' }  \left( \varphi, \varphi' \right) 
	= \frac{V}{2} \left[  1 + \eta \eta' e^{ - i \left( \varphi - \varphi' \right) } \right] \,.
\end{align}
The same limit of Eq.~\eqref{eq:chiral_disorder_s} turns Eq.~\eqref{V2} into Eq.~\eqref{eq:V_tilde} of the main text.

\subsection{Modified scattering vertex}\label{app:disorder_mod}
The modified scattering vertex \( { \left[ V_0 \right] }_{\lambda \eta } \) is given by Eq.~\eqref{eq:V0}. For intra-band pairing only, Eq.~\eqref{eq:V_tilde} is reduced to 
\begin{align}
    {}^{ \lambda }_{ \lambda }{\tilde{V}}^{ \eta }_{ \eta } \left( \varphi - \varphi' \right)
        = \frac{V^2}{4} \left[  1 + \lambda \eta e^{ - i \left( \varphi - \varphi' \right) } \right]
        \left[  1 + \lambda \eta e^{ - i \left( \varphi - \varphi' \right) } \right]
\end{align}
and multiplying Eq.~\eqref{eq:V0} by \( e^{ i \left( \varphi - \varphi' \right) } \) we obtain
\begin{align}
    { \left[ V_0 \right] }_{\lambda \eta }
        & = e^{ i \left( \varphi - \varphi' \right) } {}^{ \lambda }_{ \lambda }{\tilde{V}}^{ \eta }_{ \eta } \left( \varphi - \varphi' \right)
        \notag \\
        & = \frac{V^2}{4} \left[  e^{ i \left( \varphi - \varphi' \right) } + \lambda \eta \right]
        \left[  1 + \lambda \eta e^{ - i \left( \varphi - \varphi' \right) } \right]
        \notag \\
        & =\frac{V^2}{4} \left[ e^{ i \left( \varphi - \varphi' \right) } + e^{ - i \left( \varphi - \varphi' \right) } + 2 \lambda \eta \right] \,,
\end{align}
which is simplified to the form of Eq.~\eqref{eq:V0_a}.
\section{The contribution of the inter-band Cooper pairs}
\label{app:inter-band_pairing}

In this appendix we show that the contribution of the terms with $\eta \neq \eta'$ to the integral equation \eqref{eq:Cooperon} is negligible in the range of parameters specified by Eq.~\eqref{eq:range}.
To this end we compute $\overline{ \mathbb{G} }_{ + }^{ - }$ setting $\nu_+ = \nu_- = \nu_0$, i.e. we set $x=0$.
Using the definition, Eq.~\eqref{eq:Q} the simple calculation gives ($\varepsilon_n >0$)
\begin{align}\label{GG_pm}
	\overline{ \mathbb{G} }^+_- \left( \varphi, \mathbf{q}, \mathbf{B}, \varepsilon_n \right) 
	\!\approx \!  
	\frac{2 \pi \nu_0}
	{
	 2  \varepsilon_n  
    + \Delta_{\mathrm{SO}}
	+ i v_F ( q_x + q_y)
	+ \frac{ 1 }{ \tau }
	}.
\end{align}
In contrast to Eq.~\eqref{eq:GG_bar}, Eq.~\eqref{GG_pm} does not contain the field $\mathbf{B}$.
Therefore, at least in the limit $x=0$ it contributes nothing both to the Lifhsitz invariant and to the PHE.
The field dependence might appear in  Eq.~\eqref{GG_pm} at higher orders in $x$.
This ensures that the contribution of inter-band processes to PHE is negligible.
The same is true for the Lifshitz invariant since Eq.~\eqref{GG_pm} contains an additional parameter, $\max\{ T_c, 1/\tau \} / \Delta_{\mathrm{SO}}$ that is small in the regime, Eq.~\eqref{eq:range}.
We conclude that the contributions of the inter-band Cooper pairs are negligible in the regime, Eq.~\eqref{eq:range}.

\begin{widetext}
\section{Detailed derivation of the Cooper pair dispersion relation}

In the following sections we provide a detailed derivation of the calculation of the Cooper pair dispersion relation described in Sec.~\ref{sec:calculation}. First, in Appendix~\ref{app:Cooperon_matrix}, we detail how the Cooperon vertex's Dyson equation~\eqref{eq:Cooperon_b} is reformulated into the matrix notation of Eqs.~\eqref{eq:C_vec}-\eqref{eq:A_def}. 
Next, in Appendix~\ref{app:Cooperon_solution}, we solve the Cooperon vertex equation by expanding to second order in both \(q v_F/ \max\{T_c, 1/\tau\}, B/ \max\{T_c, 1/\tau\} \ll 1 \), obtaining an explicit solution for the Coopron vertex \( \Gamma_\lambda \left( \varphi \right) \) [Eq.~\eqref{eq:Gamma_harmonics}]. 
Lastly, in Appendix~\ref{app:polarization_bubble}, we substitute \( \Gamma_\lambda \left( \varphi \right) \) into the polarization operator~\eqref{eq:Pi_end}, and expand again in the same small parameters to find the pair dispersion relation \( \varepsilon \left( {\mathbf{q}} \right) \) using Eq.~\eqref{eq:pair_propagator}.

\subsection{Matrix formulation of the Copperon vertex equation}\label{app:Cooperon_matrix}

In this section we provide a detailed derivation of the matrix form of the Cooperon vertex. Substituting Eq.~\eqref{eq:Gamma_harmonics} into Eq.~\eqref{eq:Cooperon_b} we have two equations for the \( \lambda = \pm 1 \) vertices,

\begin{align}
    C_{ \lambda } ^{ \left( 0 \right) } 
    + C_{ \lambda } ^{ \left( c \right) } \cos \varphi 
    + C_{ \lambda } ^{ \left( s \right) } \sin  \varphi
        = \lambda & + 
        \int_{ 0 }^{ 2 \pi } \frac{d \varphi'}{2 \pi}
        { \left[ V_0 \right] }_{ \lambda \lambda } \left( \varphi - \varphi' \right)
        \overline{\mathbb{G}}^{ \lambda }_{ \lambda } \left( \varphi' \right)
        \left[ 
            C_{ \lambda } ^{ \left( 0 \right) } 
            + C_{ \lambda } ^{ \left( c \right) } \cos \varphi 
            + C_{ \lambda } ^{ \left( s \right) } \sin  \varphi
        \right]
    \notag \\
     & + 
        \int_{ 0 }^{ 2 \pi } \frac{d \varphi'}{2 \pi}
        { \left[ V_0 \right] }_{ \lambda, -\lambda} \left( \varphi - \varphi' \right)
        \overline{\mathbb{G}}^{ -\lambda }_{ -\lambda } \left( \varphi' \right)
        \left[ 
            C_{ -\lambda } ^{ \left( 0 \right) } 
            + C_{ -\lambda } ^{ \left( c \right) } \cos \varphi 
            + C_{ -\lambda } ^{ \left( s \right) } \sin  \varphi
        \right]
       \notag \\
            = \lambda & + 
        \int_{ 0 }^{ 2 \pi } \frac{d \varphi'}{2 \pi}
        \frac{V^2}{2} \left[\cos\left( \varphi - \varphi' \right) + 1 \right]
        \overline{\mathbb{G}}^{ \lambda }_{ \lambda } \left( \varphi' \right)
        \left[ 
            C_{ \lambda } ^{ \left( 0 \right) } 
            + C_{ \lambda } ^{ \left( c \right) } \cos \varphi 
            + C_{ \lambda } ^{ \left( s \right) } \sin  \varphi
        \right]
    \notag \\
     & + 
        \int_{ 0 }^{ 2 \pi } \frac{d \varphi'}{2 \pi}
        \frac{V^2}{2} \left[\cos\left( \varphi - \varphi' \right) - 1 \right]
        \overline{\mathbb{G}}^{ -\lambda }_{ -\lambda } \left( \varphi' \right)
        \left[ 
            C_{ -\lambda } ^{ \left( 0 \right) } 
            + C_{ -\lambda } ^{ \left( c \right) } \cos \varphi 
            + C_{ - \lambda } ^{ \left( s \right) } \sin  \varphi
        \right]
    \,,
\end{align}
where the explicit expression of \( { \left[ V_0 \right] }_{\lambda\eta} \), Eq.~\eqref{eq:V0_a}, is substituted in. 
Using the identity \( \cos( \varphi - \varphi' ) = \cos( \varphi ) \cos( \varphi' ) + \sin( \varphi ) \sin( \varphi' ) \) we equate the coefficient's \( \varphi\)-independent, \( \propto \cos(\varphi) \) and \( \propto \sin(\varphi) \) terms for each of the two bands. This results in six coupled equations for the coefficients of the Cooperon vertex \eqref{eq:Gamma_harmonics}:
\begin{subequations}
    \begin{align}
        C_{ \lambda } ^{ \left( 0 \right) }
    		=  
    		\lambda
    		+ \frac{V^2}{4\pi} \int_{ 0 }^{ 2 \pi } \dd{ \varphi' }
    		& \Bigg\{
    		\overline{\mathbb{G}}^{ \lambda }_{ \lambda } \left( \varphi' \right)
    		\left(
    		      C_{ \lambda } ^{ \left( 0 \right) } 
    		      + C_{ \lambda } ^{ \left( c \right) } \cos{\varphi'}
    		      + C_{ \lambda } ^{ \left( s \right) } \sin{\varphi'}
    		\right)
    		\notag\\
    		&
    		-
    		\overline{\mathbb{G}}^{ -\lambda }_{ -\lambda } \left( \varphi' \right)
    		\left(
    		      C_{ -\lambda } ^{ \left( 0 \right) } 
    		      + C_{ -\lambda } ^{ \left( c \right) } \cos{\varphi'}
    		      + C_{ -\lambda } ^{ \left( s \right) } \sin{\varphi'}
    		\right)
    		\Bigg\}
        \,,
\\
        C_{ \lambda } ^{ \left( c \right) }
    		=  
    		\hphantom{ 1 + }
    		\frac{V^2}{4\pi} \int_{ 0 }^{ 2 \pi } \dd{ \varphi' }
    		& \Bigg\{
    		\overline{\mathbb{G}}^{ \lambda }_{ \lambda } \left( \varphi' \right)
            \cos{\varphi'}
    		\left(
                C_{ \lambda } ^{ \left( 0 \right) } 
    	        + C_{ \lambda } ^{ \left( c \right) } \cos{\varphi'}
    		      + C_{ \lambda } ^{ \left( s \right) } \sin{\varphi'}
    		\right)
    		\notag\\
    		&
    		+
    		\overline{\mathbb{G}}^{ -\lambda }_{ -\lambda } \left( \varphi' \right)
    		\cos{\varphi'}
            \left(
        	    C_{ -\lambda } ^{ \left( 0 \right) } 
        		+ C_{ -\lambda } ^{ \left( c \right) } \cos{\varphi'}
        		+ C_{ -\lambda } ^{ \left( s \right) } \sin{\varphi'}
    		\right)
    		\Bigg\}
        \,,
\\
        C_{ \lambda } ^{ \left( s \right) }
    		=  
    		\hphantom{ 1 + }
    		\frac{V^2}{4\pi} \int_{ 0 }^{ 2 \pi } \dd{ \varphi' }
    		& \Bigg\{
    		\overline{\mathbb{G}}^{ \lambda }_{ \lambda } \left( \varphi' \right)
            \sin{\varphi'}
    		\left(
                C_{ \lambda } ^{ \left( 0 \right) } 
    	        + C_{ \lambda } ^{ \left( c \right) } \cos{\varphi'}
    		      + C_{ \lambda } ^{ \left( s \right) } \sin{\varphi'}
    		\right)
    		\notag\\
    		&
    		+
    		\overline{\mathbb{G}}^{ -\lambda }_{ -\lambda } \left( \varphi' \right)
    		\sin{\varphi'}
            \left(
        	    C_{ -\lambda } ^{ \left( 0 \right) } 
        		+ C_{ -\lambda } ^{ \left( c \right) } \cos{\varphi'}
        		+ C_{ -\lambda } ^{ \left( s \right) } \sin{\varphi'}
    		\right)
    		\Bigg\}
    \,.
    \end{align}
\end{subequations}

As there are many similar angular integrals we employ the shorthand of Eq.~\eqref{eq:A_def} and write these six equations as
\begin{subequations}\label{eq:C_terms}
	\begin{align}
		C_{ \lambda }^{ \left( 0 \right) }
		= &
		      \lambda
    		+
    		A_\lambda^{00} C_{ \lambda }^{ \left( 0 \right) }
    		+ 
    		A_\lambda^{10} C_{ \lambda }^{ \left( c \right) }
    		+
    		A_\lambda^{01} C_{ \lambda }^{ \left( s \right) }
    		-
    		A_{-\lambda}^{00} C_{ -\lambda } ^{ \left( 0 \right) }
    		- 
    		A_{-\lambda}^{10} C_{ -\lambda }^{ \left( c \right) }
    		-
    		A_{-\lambda}^{01} C_{ -\lambda }^{ \left( s \right) }
        \,,
		\\
		C_{ \lambda }^{ \left( c \right) }
    		= &
            \hphantom{ 11 + }
    		A_\lambda^{10} C_{ \lambda }^{ \left( 0 \right) }
    		+ 
    		A_\lambda^{20} C_{ \lambda }^{ \left( c \right) }
    		+
    		A_\lambda^{11} C_{ \lambda }^{ \left( s \right) }
    		+
    		A_{-\lambda}^{10} C_{ -\lambda } ^{ \left( 0 \right) }
    		+ 
    		A_{-\lambda}^{20} C_{ -\lambda }^{ \left( c \right) }
    		+
    		A_{-\lambda}^{11} C_{ -\lambda }^{ \left( s \right) }
        \,, 
		\\
		C_{ \lambda }^{ \left( s \right) }
    		= &
    		\hphantom{ 11 + }
    		A_\lambda^{01} C_{ \lambda }^{ \left( 0 \right) }
    		+ 
    		A_\lambda^{11} C_{ \lambda }^{ \left( c \right) }
    		+
    		A_\lambda^{02} C_{ \lambda }^{ \left( s \right) }
    		+
    		A_{-\lambda}^{01} C_{ -\lambda } ^{ \left( 0 \right) }
    		+ 
    		A_{-\lambda}^{11} C_{ -\lambda }^{ \left( c \right) }
    		+
    		A_{-\lambda}^{02} C_{ -\lambda }^{ \left( s \right) }
        \,.
	\end{align}
\end{subequations}
Naturally these equations lend themselves nicely to a matrix representation with the coefficients \( C_\lambda^{\left( i \right) } \) arranged in a (six-entry) column vector \( \vb{C}\) [Eq.~\eqref{eq:C_vec}]. 
Equation.~\eqref{eq:C_terms} reads as \(
    \vb{C} = \left[- 1,  0,  0,  1,  0,   0 \right]^{\mathrm{t}} + \mathbb{M}  \vb{C} \,,
\)
with the \(6\times6\) matrix \( \mathbb{M} \) defined in Eq.~\eqref{eq:M_matrix}.

\subsection{Solving \texorpdfstring{Eq.~\eqref{eq:M}}{Eq.53} for the Cooperon vertex \texorpdfstring{\(\vb{C}\)}{C} }\label{app:Cooperon_solution}

In this section we provide the details required for solving Eq.~\eqref{eq:M} to find the Cooperon vertex \( \vb{C} \) [Eq.~\eqref{eq:C_vec}].
To solve Eq.~\eqref{eq:M} we start by calculating the 12 possible values of the \( A_\lambda^{ m n } \) [Eq.~\eqref{eq:A_def}], which are the entries of the matrix \( \mathbb{M} \) [Eq.~\eqref{eq:M_matrix}].
Each is calculated by substituting the expression for the Green's function pair \( \overline{ \mathbb{G} }^{ \lambda }_{ \lambda } \), Eq.~\eqref{eq:GG_bar}, into Eq.~\eqref{eq:A_def}, then expanding the expression to second order in both \(q v_F/ \max\{T_c, 1/\tau\}\) and \(B/ \max\{T_c, 1/\tau\}\) and preforming the angular integral. This results in the following expressions which employs the \(  x, y \) component of \( \mathbf{Q}^\lambda \) [Eq.~\eqref{eq:Q}] up to fourth order [the expansion order is discussed in the main text following Eq.\eqref{eq:Q}]:
\begin{subequations}\label{eq:As}
	\begin{align}
		A_\lambda^{00}
		      & = \frac{V^2}{4 \pi} \int_{ 0 }^{ 2\pi } \dd{ \varphi' } 
		      \overline{ \mathbb{G} }^\lambda_\lambda \left( \varphi' \right)
    		  = \frac{\nu_\lambda}{\nu_0} \tau^{-1} 
    		\left\{
    		\frac{1}{ 2 \left( \tau^{-1} + \varepsilon_n \right) }
    		-
    		v_F^2 \frac{ \left( Q_x^\lambda \right)^2 + \left( Q_y^\lambda \right)^2 }{ \left( \tau^{-1} + \varepsilon_n \right)^3 }
    		+
    		3 v_F^4 \frac{ \left[ \left( Q_x^\lambda \right)^2 + \left( Q_y^\lambda \right)^2 \right]^2 }{ \left( \tau^{-1} + \varepsilon_n \right)^5 }
    		\right\}
    \\
        A_\lambda^{10}
    		& = \frac{ V^2 }{ 4 \pi} \int_{ 0 }^{ 2\pi } \dd{ \varphi' } 
    		\overline{ \mathbb{G} }^\lambda_\lambda \left( \varphi' \right)
    		\cos( \varphi' )
            = - i \frac{\nu_\lambda}{\nu_0}  \tau^{-1} 
    		\left\{
    		\frac{ Q_x^\lambda v_F }{ 2 \left( \tau^{-1} + \varepsilon_n \right)^2 }
    		-
    		3 Q_x^\lambda v_F^3 \frac{ \left( Q_x^\lambda \right)^2 + \left( Q_y^\lambda \right)^2 }{ 2 \left( \tau^{-1} + \varepsilon_n \right)^4 }
    		\right\}				
    \\
    	A_\lambda^{01}
    		& = \frac{V^2}{4 \pi} \int_{ 0 }^{ 2\pi } \dd{ \varphi' } 
    		\overline{ \mathbb{G} }^\lambda_\lambda \left( \varphi' \right)
    		\sin[]( \varphi' )
    	    = - i \frac{\nu_\lambda}{\nu_0}  \tau^{-1} 
    		\left\{
    		\frac{ Q_y^\lambda v_F }{ 2 \left( \tau^{-1} + \varepsilon_n \right)^2 }
    		-
    		3 Q_y^\lambda v_F^3 \frac{ \left( Q_x^\lambda \right)^2 + \left( Q_y^\lambda \right)^2 }{ 2 \left( \tau^{-1} + \varepsilon_n \right)^4 }
    		\right\}				
	\\
	    A_\lambda^{11}
    		& = \frac{V^2}{ 4 \pi} \int_{ 0 }^{ 2\pi } \dd{ \varphi' } 
    		\overline{ \mathbb{G} }^\lambda_\lambda \left( \varphi' \right)
    		\cos[]( \varphi' ) \sin[]( \varphi' )
    	    = - \frac{\nu_\lambda}{\nu_0} \tau^{-1} 
    		\left\{
    		\frac{ Q_x^\lambda Q_y^\lambda v_F^2 }{ 2 \left( \tau^{-1} + \varepsilon_n \right)^3 }
    		-
    		2 Q_x^\lambda Q_y^\lambda v_F^4 \frac{ \left( Q_x^\lambda \right)^2 + \left( Q_y^\lambda \right)^2 }{ \left( \tau^{-1} + \varepsilon_n \right)^4 }
    		\right\}				
	\\
	    A_\lambda^{20} 
    		& = \frac{V^2}{4 \pi} \int_{ 0 }^{ 2\pi } \dd{ \varphi' } 
    		\overline{ \mathbb{G} }^\lambda_\lambda \left( \varphi' \right)
    		\cos[2]( \varphi' )
    		\nonumber \\
    		& = \frac{\nu_\lambda}{\nu_0} \tau^{-1} 
    		\left\{
    		\frac{1}{ 4 \left( \tau^{-1} + \varepsilon_n \right) }
    		-
    		v_F^2 \frac{ 3 \left( Q_x^\lambda \right)^2 + \left( Q_y^\lambda \right)^2 }{ 4 \left( \tau^{-1} + \varepsilon_n \right)^3 }
    		+
    		v_F^4 \frac{ \left[ 5 \left( Q_x^\lambda \right)^4 + 6 \left( Q_x^\lambda \right)^2 \left( Q_y^\lambda \right)^2 + \left( Q_y^\lambda \right)^4 \right]^2 }{ 2 \left( \tau^{-1} + \varepsilon_n \right)^5 }
    		\right\}
	\\		
	    A_\lambda^{02} 
    		& = \frac{V^2}{4 \pi} \int_{ 0 }^{ 2\pi } \dd{ \varphi' } 
    		\overline{ \mathbb{G} }^\lambda_\lambda \left( \varphi' \right)
    		\sin[2]( \varphi' )
    		\nonumber \\
    		& = \frac{\nu_\lambda}{\nu_0} \tau^{-1} 
    		\left\{
    		\frac{1}{ 4 \left( \tau^{-1} + \varepsilon_n \right) }
    		-
    		v_F^2 \frac{ \left( Q_x^\lambda \right)^2 + 3 \left( Q_y^\lambda \right)^2 }{ 4 \left( \tau^{-1} + \varepsilon_n \right)^3 }
    		+
    		v_F^4 \frac{ \left[ \left( Q_x^\lambda \right)^4 + 6 \left( Q_x^\lambda \right)^2 \left( Q_y^\lambda \right)^2 + 5 \left( Q_y^\lambda \right)^4 \right]^2 }{ 2 \left( \tau^{-1} + \varepsilon_n \right)^5 }
    		\right\}
    \,.
	\end{align}
\end{subequations}
Equations~\eqref{eq:As} expresses Eq.~\eqref{eq:M_matrix} fully.

Next, we invert \( \check{\mathbb{M}} \equiv \left( \mathbb{M} - \mathbb{1}_6 \right) \) which appears in Eq.~\eqref{eq:M}. 
The inverse matrix entries all have the same general form,
\begin{align}\label{eq:M-1}
	\left[ \check{ \mathbb{M} }^{-1} \right]_{ij}
	= \frac{ \mathit{m}_{ij} }{ \det[ \check{ \mathbb{M} } ] }
    \,,
\end{align}
where \( \mathit{m}_{ij} \) is the numerator of the \( ij \)-entry (comprised of the minors of the original matrix) and the denominator is the determinant of the original matrix, identical for all entries.

Each \( \left[ \check{ \mathbb{M} }^{-1} \right]_{ij} \) is a cumbersome expression that can effectively be dealt with using \textit{Mathematica}.
Here we present only a single entry for illustration purposes.
The numerator is expressed using \(  A_\lambda^{ij} \) of Eq.~\eqref{eq:A_def},
\begin{align}\label{}
	m_{1,1} = 
    	& A_-^{00} \left[ A_+^{20} ( A_-^{02}+ A_+^{02}-1)-( A_-^{11}+ A_+^{11})^2+( A_+^{20}-1) ( A_-^{02}+ A_+^{02}-1) \right] - \left[ \left(  A_-^{10} \right)^2 ( A_-^{02}+ A_+^{02}-1) \right]
       	\notag \\
    	& +  A_-^{10} ( A_-^{11}+ A_+^{1}) (2  A_-^{01}- A_+^{01})+ A_-^{10} A_+^{10} ( A_-^{02}+A_+^{02}-1) - A_+^{20} \left(  A_-^{01} \right)^2 + A_+^{20} A_-^{01}  A_+^{01}- A_+^{20}  A_-^{02}
       	\notag \\
    	& +  A_-^{10} ( A_-^{11}+ A_+^{1}) (2  A_-^{01}- A_+^{01})+ A_-^{10} A_+^{10} ( A_-^{02}+A_+^{02}-1) - A_+^{20} \left(  A_-^{01} \right)^2 + A_+^{20} A_-^{01}  A_+^{01}- A_+^{20}  A_-^{02}
       	\notag \\
    	& - A_+^{02} ( A_+^{20}+ A_+^{20})+ A_+^{20} + \left(  A_-^{11} \right)^2-  A_-^{11}  A_-^{01} \left(  A_+^{10} \right) + 2  A_-^{11} A_+^{11}-\left(  A_-^{01} \right)^2 A_+^{20} + \left(  A_-^{01} \right)^2
        \notag \\
    	& - A_-^{01} A_+^{10} A_+^{1}+ A_-^{01} A_+^{20}  A_+^{01}- A_-^{01}  A_+^{01}- A_-^{02} A_+^{20}+ A_-^{02} + A_+^{20} + \left( A_+^{11} \right)^2+\ A_+^{02}-1
     \,.
\end{align}
The rest of the numerators have similar expressions, and we expand these to fourth order in \( \vb{Q}^\lambda \).

The determinant is given by a similar expression containing \(  A_\lambda^{ij} \). Keeping terms up to fourth order in \( \vb{Q}^\lambda \) we have 
\begin{align}\label{detM}
	\det[\check{\mathbb{M}}] = 
    	\frac{ \tau \varepsilon_n }{2}
    	+
    	& \frac{\tau^2 v_F^2}{32}
    	\left[
    	\left( 3 \left( Q_x^- \right)^2 -2 Q_x^- Q_x^+ + 3\left( Q_x^+ \right)^2 \right) 
    	+
    	\left( 3 \left( Q_y^- \right)^2 -2 Q_y^- Q_y^+ + 3\left( Q_y^+ \right)^2 \right)
    	\right]
    	\notag \\
    	- \frac{\tau^4 v_F^4}{256}
    	& \left[
    	- 15 \left( Q_x^- \right)^4 
    	+ 10 \left( Q_x^- \right)^3 Q_x^+ 
    	+ 18 \left( Q_x^- \right)^2 \left( Q_x^+ \right)^2 
    	+ 10 Q_x^- \left( Q_x^+ \right)^3 
    	- 15 \left( Q_x^+ \right)^4 
    	\right.
    	\notag \\
    	& \left.
    	- 30 \left( Q_x^- \right)^2 \left( Q_y^- \right)^2 
    	+ 10 Q_x^- Q_x^+ \left( Q_y^- \right)^2 
    	+ 22 \left( Q_x^+ \right)^2 \left( Q_y^- \right)^2 
    	- 15 \left( Q_y^- \right)^4
    	\right.
    	\notag \\
    	& \left.
    	+ 10 \left( Q_x^- \right)^2 Q_y^- Q_y^+ 
    	- 8 Q_x^- Q_x^+ Q_y^- Q_y^+ 
    	+ 10 \left( Q_x^+ \right)^2 Q_y^- Q_y^+ 
    	+ 10 \left( Q_y^- \right)^3 Q_y^+ 
    	+ 22 \left( Q_x^- \right)^2 \left( Q_y^+ \right)^2 
    	\right.
    	\notag \\
    	& \left.
    	+ 10 Q_x^- Q_x^+ \left( Q_y^+ \right)^2
    	- 30 \left( Q_x^+ \right)^2 \left( Q_y^+ \right)^2
    	+ 18 \left( Q_y^- \right)^2 \left( Q_y^+ \right)^2 
    	+ 10 Q_y^- \left( Q_y^+ \right)^3 
    	- 15 \left( Q_y^+ \right)^4 
    	\right]
    	\,.
\end{align}

With the entries of the inverse matrix \( \mathbb{M}^{-1} \), Eq.~\eqref{eq:M-1}, the components of Eq.~\eqref{eq:C_vec} are given by
\begin{subequations}
	\begin{align}
		C_{ + } ^{ \left( 0 \right) }
		& = - \left[ \check{ \mathbb{M} }^{-1} \right]_{11} + \left[ \check{ \mathbb{M} }^{-1} \right]_{14}
        \,,
		\\
		C_{ + } ^{ \left( c \right) }
		& = - \left[ \check{ \mathbb{M} }^{-1} \right]_{21} + \left[ \check{ \mathbb{M} }^{-1} \right]_{24}
        \,,
		\\
		C_{ + } ^{ \left( s \right) }
		& = - \left[ \check{ \mathbb{M} }^{-1} \right]_{31} + \left[ \check{ \mathbb{M} }^{-1} \right]_{34}
        \,,
		\\
		C_{ - } ^{ \left( 0 \right) }
		& = - \left[ \check{ \mathbb{M} }^{-1} \right]_{41} + \left[ \check{ \mathbb{M} }^{-1} \right]_{44}
        \,,
		\\
		C_{ - } ^{ \left( c \right) }
		& = - \left[ \check{ \mathbb{M} }^{-1} \right]_{51} + \left[ \check{ \mathbb{M} }^{-1} \right]_{54}
        \,,
		\\
		C_{ - } ^{ \left( s \right) }
		& = - \left[ \check{ \mathbb{M} }^{-1} \right]_{61} + \left[ \check{ \mathbb{M} }^{-1} \right]_{64}
        \,,
	\end{align}
\end{subequations}
and these are substituted into Eq.~\eqref{eq:Gamma_harmonics} to give the final result of the two Cooperon vertices \( \Gamma_\pm \left( \varphi \right) \).

\subsection{Calculation of the pair dispersion from the polarization operator, \texorpdfstring{Eq.~\eqref{eq:Pi_end}}{Eq.56}}\label{app:polarization_bubble}

Here we detail the calculation the polarization operator \( \Pi \left( \vb{q} \right) \) [Eq.~\eqref{eq:Pi_end}] and the pair dispersion relation~\eqref{eq:pair_propagator}.
The dispersion relation in turn allows us to identify \( \mathcal{L}_1 \) in Eq.~\eqref{eq:L1}, \( \xi_2^2 \)  in Eq.~\eqref{eq:xi_2}, and \( \mathcal{L}_2 \) in Eq.~\eqref{eq:L2}.

The two terms in Eq.~\eqref{eq:Pi_end} are calculated by substituting the Cooperon vertex \( \Gamma_{ \lambda } \left( \varphi \right) \) [Eq.~\eqref{eq:Gamma_harmonics}] and the Green's function pair \( \mathbb{G}_{ \lambda }^{ \lambda } \) [Eq.~\eqref{eq:GG_bar}] and expanding to fourth order \( \vb{Q}^\lambda\) [Eq.~\eqref{eq:Q}]. The appropriate expansion order is discussed in the paragraph following Eq.~\eqref{eq:Q}. After the angular integration, Eq.~\eqref{eq:Pi_end} takes the form
\begin{align}\label{eq:bubble_short}
	\nu_0^{-1} \Pi \left( \vb{q} \right)
    &  = 
		\left[
		T \sum_{ \varepsilon_n > 0 } \frac{ 2 \pi }{ \varepsilon_n }
		-
		\left(
		\Xi_B
		+ \xi_x q_x 
		+ \xi_y  q_y
		+ \xi_{ 0 }^2 \vb{q}^2
		+ \xi_{ x^2 }^2  q_x^2
		+ \xi_{ y^2 }^2   q_y^2
		+ \xi_{ x y }^2 q_x  q_y
		\right) 
		\right]
    \,,
\end{align}
with expansion coefficients proportional to the components of the pair momentum \( \vb{q} \) up to second order.
Each of these coefficients contains sums over Matsubara frequencies,
\begin{subequations}
    \begin{align}
        \Xi_B 
    		= &  T \sum_{ \varepsilon_n > 0 }  \frac{ 4 \pi \tau B^2 }{ \varepsilon_n^2 \left( 1 + 4 \tau \varepsilon_n \right) }
        \,,
        \\
        \xi_{ 0 }^2
    		= & T \sum_{ \varepsilon_n > 0 } \frac{ \pi \tau v_F^2 }{ 2 \varepsilon_n^2 \left( 1 + 2 \tau \varepsilon_n \right) }
        \,,
        \\
        \xi_x
		      = & + x B_y T \sum_{ \varepsilon_n > 0 } \frac{ \pi \tau v_F }{ \varepsilon_n^2 \left( 1 + 4 \tau \varepsilon_n \right) }
        \,,
        \\
        \xi_y
		      = & - x B_x T \sum_{ \varepsilon_n > 0 } \frac{ \pi \tau v_F }{ \varepsilon_n^2 \left( 1 + 4 \tau \varepsilon_n \right) }
        \,,
        \\
        \xi_{xy}^2
		    = & 2 B_x B_y \pi
        		T \sum_{ \varepsilon_n > 0 } 
        		\frac{ \tau^3 v_F^2 \left( 13 + 72 \tau \varepsilon_n +  96 \tau^2 \varepsilon_n^2 \right)  }{ \varepsilon_n^2 \left( 1 + 2 \tau \varepsilon_n \right)^3 \left( 1 + 4 \tau \varepsilon_n \right)^2 }
        \,,
        \\
        \xi_{ x^2 }^2
		      = & - B^2 \pi
		      T \sum_{ \varepsilon_n > 0 } 
        		\frac{ \tau^2 v_F^2 \left( 4 + 37 \tau \varepsilon_n + 104 \tau^2 \varepsilon_n^2 + 96 \tau^3 \varepsilon_n^3 \right)  }{ 2 \varepsilon_n^3 \left( 1 + 2 \tau \varepsilon_n \right)^3 \left( 1 + 4 \tau \varepsilon_n \right)^2 }
		       - B_y^2
		      T \sum_{ \varepsilon_n > 0 } 
        		\frac{ \tau^3 v_F^2 \left( 13 + 72 \tau \varepsilon_n + 96 \tau^2 \varepsilon_n^2 \right)  }{ \varepsilon_n^2 \left( 1 + 2 \tau \varepsilon_n \right)^3 \left( 1 + 4 \tau \varepsilon_n \right)^2 }
        \,,
      \\
      	\xi_{ y^2 }^2
    		= & - B^2 \pi
    		T \sum_{ \varepsilon_n > 0 } 
        		\frac{ \tau^2 v_F^2 \left( 4 + 37 \tau \varepsilon_n + 104 \tau^2 \varepsilon_n^2 + 96 \tau^3 \varepsilon_n^3 \right)  }{ 2 \varepsilon_n^3 \left( 1 + 2 \tau \varepsilon_n \right)^3 \left( 1 + 4 \tau \varepsilon_n \right)^2 }
             - B_x^2
    		T \sum_{ \varepsilon_n > 0 } 
        		\frac{ \tau^3 v_F^2 \left( 13 + 72 \tau \varepsilon_n + 96 \tau^2 \varepsilon_n^2 \right)  }{ \varepsilon_n^2 \left( 1 + 2 \tau \varepsilon_n \right)^3 \left( 1 + 4 \tau \varepsilon_n \right)^2 }
        \,.
    \end{align}
\end{subequations}

The repeating sums lend themselves to the definition of Matsubara sums in Eq.~\eqref{eq:notation_sum} making the expression compact,
\begin{subequations}\label{eq:disp_coeff}
	\begin{align}
		\Xi_B = & B^2 4 \tau \gamma_2
    \,,
    \label{eq:Xi_B}
		\\
		\xi_0 ^2 = & \tau v_F^2 \frac{\gamma_1}{2}
    \,,
		\\
		\xi_x = & + B_y \left( \tau  v_F \gamma_2 \right) x
    \,,
    \label{eq:xi_x}
		\\
		\xi_y = & - B_x \left( \tau  v_F\gamma_2 \right) x
    \,,
    \label{eq:xi_y}
		\\
		\xi_{xy}^2 = & B_x B_y 
            2 \pi \tau^2 v_F^2  
		      \left( 13 \tau u_2 + 72 \tau^2 u_1 + 96 \tau^3 u_0 \right)
    \,,
    \label{eq:xi_xy}
		\\
		\xi_{x^2}^2 = & - B_y^2 \pi \tau^2 v_F^2 \left( 13 \tau u_2 + 72 \tau^2 u_1 + 96 \tau^3 u_0 \right) - B^2 \frac{\pi}{2} \tau^2 v_F^2 \left( 4 u_3 + 37 \tau u_2 + 104 \tau^2 u_1 + 96 \tau^3 u_0 \right)
    \,,
    \label{eq:xi_x2}
		\\
		\xi_{y^2}^2 = & - B_x^2 \pi \tau^2 v_F^2 \left( 13 \tau u_2 + 72 \tau^2 u_1 + 96 \tau^3 u_0 \right) - B^2 \frac{\pi}{2} \tau^2 v_F^2 \left( 4 u_3 + 37 \tau u_2 + 104 \tau^2 u_1 + 96 \tau^3 u_0 \right)
    \label{eq:xi_y2}
	\,.
	\end{align}
\end{subequations}
In order to better see the algebraic structure due to the system's symmetry, we simplify Eqs.~\eqref{eq:xi_x2} and \eqref{eq:xi_y2} by employing the definitions of \( \rho_j \), Eq.~\eqref{eq:rhos}. This gives \(  \xi_{x^2}^2 =  - \frac{1}{2} B_y^2 \rho_1 - B^2 \rho_2 ,\, \xi_{y^2}^2 = - \frac{1}{2} B_x^2 \rho_1 - B^2 \rho_2 \). 
Equation~\eqref{eq:Xi_B} modifies the critical temperature according to Eq.~\eqref{eq:TcR}.

\subsection{Symmetry-respecting form for the pair dispersion}\label{app:L}

With \( \Pi \left( \vb{q} \right) \), Eq.~\eqref{eq:bubble_short}, known, the pair dispersion \( \varepsilon \left( {\mathbf{q}} \right)  \) is given by Eq.~\eqref{eq:pair_propagator}. Our remaining task is to write these expressions in a form which better represents the symmetry of the problem, thus identifying \( \mathcal{L}_1 \) and \( \mathcal{L}_2 \) in Eq.~\eqref{eq:dispersion}. The linear terms in the field, Eqs.~\eqref{eq:xi_x} and \eqref{eq:xi_y}, are
\begin{align}
    \xi_x q_x + \xi_y q_y 
        = \left( B_y q_x - B_x q_y \right) \left( \tau v_F \gamma_2 \right) x 
        = \left( \tau v_F \gamma_2 \right) x \left( \vb{q} \cross \vb{B} \right) \cdot \hat{z}
\end{align}
which according to Eq.~\eqref{eq:dispersion} gives the expression of \( \mathcal{L}_1 \) in Eq.~\eqref{eq:L1}.

The terms quadratic in the field, Eqs.~\eqref{eq:xi_xy}, \eqref{eq:xi_x2} and \eqref{eq:xi_y2},
\begin{align}\label{}
	\xi_{x^2}^2 q_x^2 + \xi_{xy}^2 q_x q_y + \xi_{y^2}^2 q_y^2
        & = \left( - \frac{1}{2} B_y^2 \rho_1 - B^2 \rho_2 \right) q_x^2
        + B_x B_y \rho_1 q_x q_y
        + \left( - \frac{1}{2} B_x^2 \rho_1 - B^2 \rho_2 \right) q_y^2
    \notag \\
        & = - \frac{1}{2} \rho_1 \left( B_y^2 q_x^2 - 2 B_x B_y q_x q_y + B_x^2 q_y^2 \right)
        - \rho_2 B^2 \left( q_x^2 + q_y^2 \right)
    \notag \\
        & = \left( - \rho_2 - \frac{\rho_1}{4} \right) q^2 B^2 - \frac{\rho_1}{4} \left[ \left( \vb{q} \cdot \vb{B} \right)^2 - \left( \vb{q} \cross \vb{B} \right)^2 \right]
    \notag \\
        & \equiv \xi_2^2 q^2 B^2 - \mathcal{L}_2 \left[ \left( \vb{q} \cdot \vb{B} \right)^2 - \left( \vb{q} \cross \vb{B} \right)^2 \right]
    \,,
\end{align}
allow us to identify \( \xi_2^2 \)  of Eq.~\eqref{eq:xi_2} and \( \mathcal{L}_2 \) of Eq.~\eqref{eq:L2}.

\section{Contributions to the dispersion \texorpdfstring{\eqref{eq:dispersion}}{(3)} due to the field dependence of the pairing vertices}\label{app:pairing_high}

In this appendix we explicitly derive the addition to \( \mathcal{L}_1 \) and \( \mathcal{L}_2\) which arise from magnetic field related corrections to the pairing Hamiltonian, Eqs.~\eqref{eq:Hps} and \eqref{eq:Hpt}, for the singlet and triplet pairing respectively.
These corrections, \( \delta \mathcal{L}_1\) and \( \delta \mathcal{L}_2 \), are Eqs.~\eqref{eq:deltaL1}, \eqref{eq:deltaL1_triplet} and \eqref{eq:deltaL2} of the main text.

The Lifshitz invariant \( \mathcal{L}_1 \), defined by Eq.~\eqref{eq:dispersion}, is the prefactor linear in both \( \vb{q}, \vb{B} \) and thus to observe its modification due to the pairing interaction it is enough to keep terms up to second order in \( \vb{Q}^\lambda \) [Eq.~\eqref{eq:Q}]. \( \mathcal{L}_2 \) is quadratic in \( \vb{q}, \vb{B} \) and thus we need to keep fourth order in \( \vb{Q}^\lambda \). The modifications of the quadratic terms \( \delta \mathcal{L}_2 \) [Eq.~\eqref{eq:deltaL2}] are small (Eq.~\eqref{eq:ratioL2c} and discussion thereafter), and thus we only provide these correction's functional form without the details of the numerical prefactors.

\subsection{Singlet pairing interaction vertex}\label{app:pairing_s}

We start with the singlet pairing Hamiltonian~\eqref{eq:Hps}. The Hamiltonian in the basis of Eq.~\eqref{eq:psiB} is given by Eq.~\eqref{eq:Hps_chiral} with the Coopron vertices, \( \left[ i \check{\sigma}_y \right]_{\lambda_1,\lambda_2} \) and \( \left[ i \check{\sigma}_y \right]^\dagger_{\lambda_3 \lambda_4} \), given by Eqs.~\eqref{SC_Pauli_chiral} and \eqref{SC_Pauli_chiral_hc} respectively.
We expand these vertices to second order in both \(q v_F/ \max\{T_c, 1/\tau\}\) and \(B/ \max\{T_c, 1/\tau\}\). As with the entire calculation the expansion is obtained by keeping terms up to fourth order in \( \vb{Q}^\lambda \) [Eq.~\eqref{eq:Q}], yielding
\begin{subequations}
    \begin{align}\label{eq:CooperS2}
    	\left[i \check{\sigma}_y \left( \varphi, \vb{q}, \vb{B} \right) \right]
    		= & i e^{ i \varphi } \sigma_z
    		+
    		\frac{ e^{ i \varphi } }{ k_F } \left[ Q_y \cos(\varphi) - Q_x \sin(\varphi) \right] \sigma_x
    	\notag \\
    	 	& +
            \left[ 
                3 e^{ 3 i \varphi } \left( Q_x - i Q_y \right)^2 - e^{ - i \varphi} \left( Q_x + i Q_y \right)^2 - 2 e^{ i \varphi } \left( Q_x^2 + Q_y^2 \right) 
            \right]
            \frac{i \sigma_z}{8 k_F^2}
			+ S_3 \left( \vb{Q}^3 \right) \sigma_x + S_4 \left( \vb{Q}^4 \right) \sigma_z
    \end{align}
    and
    \begin{align}\label{eq:CooperS2dag}
    	\left[i \check{\sigma}_y \left( \varphi, \vb{q}, \vb{B} \right) \right]^\dagger
    	&	=  - i e^{ - i \varphi } \sigma_z
    		+
    		\frac{ e^{ - i \varphi } }{ k_F } \left[ Q_y \cos(\varphi) - Q_x \sin(\varphi) \right] \sigma_x
    	\notag \\
    	 	 + & \left[ 
                e^{ i \varphi } \left( Q_x - i Q_y \right)^2 - 3 e^{-3 i \varphi} \left( Q_x + i Q_y \right)^2 + 2 e^{ -i \varphi } \left( Q_x^2 + Q_y^2 \right) 
            \right] \frac{i \sigma_z}{8 k_F^2}
    		+ S_3^d \left( \vb{Q}^3 \right) \sigma_x 
            + S_4^d \left( \vb{Q}^4 \right) \sigma_z ,
    \end{align}
\end{subequations}
where \( S_3 \left( \vb{Q}^3 \right), S_3^d \left( \vb{Q}^3 \right), S_4 \left( \vb{Q}^4 \right),  S_4^d \left( \vb{Q}^4 \right) \) notate the third and fourth order singlet pairing vertex expansion in \( \vb{Q}^\lambda \). These cumbersome expressions are omitted from the presentation due to not contributing to the final result in the order of calculation.

We substitute Eqs.~\eqref{eq:CooperS2} and \eqref{eq:CooperS2dag} into Eq.~\eqref{eq:Bubble_w_Cooperon} to obtain the polarization operator for the singlet channel pairing \( \Pi_s \), but instead of the Cooperon ladder on the right side of Fig.~\ref{fig:Pi} we have the Cooperon vertex expanded to fourth order in \( \vb{Q}^\lambda \). The singlet polarization operator is
\begin{align}\label{eq:Pi_s_orders}
    \Pi_s
        = \Pi_s^{ \left( 0 \right) } \left( T_{c0} \right) + \Pi_s^{ \left( 2 \right) } +\Pi_s^{ \left( 4 \right) }
\end{align}
with the superscript indicating the order in \( \vb{Q}^\lambda \).
The zero order takes the well known BCS form
\begin{align}\label{eq:Cooper_log}
	\Pi_s^{ \left( 0 \right) } \left( T_{c0} \right)
		= \left[ \ln( \frac{ 2 \omega_D }{ \pi T_{c0} } e^{ \gamma_E }) \right]^{-1} \equiv \Lambda \,,
\end{align}
with \( \omega_D \) the energy upper cutoff, \( \gamma_E \approx 0.577 \) the Euler number, and defining \( \Lambda \) for notation simplification.

The second-order correction to the singlet polarization operator is
\begin{align}\label{eq:delta_L1s_calc}
	\frac{\Pi_s^{\left( 2 \right)}}{\Pi_s^{\left( 0 \right)}} 
		= - \frac{ Q_x^2 + Q_y^2 }{2 k_F^3 \Lambda } = \frac{ q_x B_y - q_y B_x }{ 2 k_F^2 \alpha \Lambda } = \frac{ \left( \vb{q} \cross \vb{B} \right) \cdot \hat{z} }{ k_F \Delta_{\mathrm{SO}} \Lambda }
\end{align}
where the SO splitting \( \Delta_{\mathrm{SO}} \), Eq.~\eqref{eq:DeltaSO}, is used.
This gives the correction \( \delta \mathcal{L}_1 \) of Eq.~\eqref{eq:deltaL1}.

The singlet fourth-order correction, i.e. \( \delta \mathcal{L}_2^s \) comes from the last term in Eq.~\eqref{eq:Pi_s_orders},
\begin{align}\label{eq:delta_L2s_calc}
	\frac{\Pi_s^{\left( 4 \right)}}{\Pi_s^{\left( 0 \right)} \left( T_{c0} \right) } 
        = \frac{1}{16} \frac{ \left( 2 q_x^2 B_y^2 + 2 q_y^2 B_x^2 - 4 q_x q_y B_x B_y \right)+ q^2 B^2 }{k_F^4 \alpha^2 \Lambda } 
	    = \frac{1}{2}
        \left[ 
            \frac{ \left( \vb{q} \cdot \vb{B} \right)^2 - \left( \vb{q} \cross \vb{B} \right)^2 }{ k_F^2 \Delta_{\mathrm{SO}}^2 \Lambda } + \frac{q^2 B^2}{2 k_F^2 \Delta_{\mathrm{SO}}^2 \Lambda }
        \right]
    	\,,
\end{align}
where the second term in the square brackets is a small correction to \( \xi_2^2 \), Eq.~\eqref{eq:xi_2}, and the first term is \( \delta \mathcal{L}_2^s \) of \eqref{eq:deltaL2}.

\subsection{Triplet channel interaction vertices}\label{app:pairing_t}

The triplet channel interaction is parametrized by the $\mathbf{d}_{\mathbf{k}}$-vector that is odd in momentum,
\begin{align}\label{eq:Hpt}
    H_{p}^t  = & \frac{g_t}{4} \sum_{\eta=x,y} \sum_{\mathbf{k},\mathbf{k}',\mathbf{q};s} 
		      \left\{ c^\dagger_{\mathbf{k}_+s_1}  
            \left[ \mathbf{d}^\eta_{\mathbf{k}} \cdot \boldsymbol{\sigma} i \sigma_y \right]_{s_1 s_2} c^\dagger_{-\mathbf{k}_-s_2}\right\}
            \left\{ c_{-\mathbf{k}'_- s_3} 
            \left[\mathbf{d}^\eta_{\mathbf{k}'} \cdot \boldsymbol{\sigma} i \sigma_y \right]^\dagger_{s_3 s_4} c_{\mathbf{k}'_+s_4} \right\}\, ,
\end{align}
where $\eta = x,y$ stands for the two components of the $\mathbf{d}_{\mathbf{k}}$-vector.

We choose the triplet interaction that is motivated by geometrical arguments. 
Such triplet interaction arises from the deformation of the spin polarization texture by the Zeeman interaction, see Fig.~\ref{fig:texture}.
The deformation at a particular momentum $\mathbf{k}$ is proportional to the cross product,
$ \boldsymbol{\gamma}(\mathbf{k}) \times \mathbf{B}$.
This observation allows one to formulate the field induced triplet order parameter for any given spin-orbit interaction \cite{Mockli2019}.
Such field induced triplets have the same transformation properties as the field that induces them.

For Rashba spin-orbit coupling, Eq.~\eqref{eq:H_SO}, writing 
$\boldsymbol{\gamma}(\mathbf{k}) \times \mathbf{B} = \alpha \hat{z} (\mathbf{B} \cdot \mathbf{k})$
we deduce the following form of the field induced triplet interactions,
\begin{align}\label{eq:d-vector}
    \mathbf{d}_\mathbf{k}^x = \hat{z} \cos \varphi \, , \quad 
    \mathbf{d}_\mathbf{k}^y = \hat{z} \sin \varphi \, ,
\end{align}
where we denote $k_x = k \cos \varphi$ and $k_y = k \sin \varphi$.
Equation~\eqref{eq:d-vector} signifies that the Zeeman field couples to $p$-wave triplet interaction channel.
It is clear that the pair of functions, Eq.~\eqref{eq:d-vector}, as well as the pair of in-plane field components, \( \mathbf{B}=(B_x,B_y) \), both transform as the $E$ irreducible representation of $C_{3 v}$ as expected.

We note in passing that the out-of-plane field belongs to the $A_2$ irreducible representation of the same group. 
The one-component triplets of the $A_2$ symmetry of the form $\mathbf{d}_\mathbf{k} = \hat{\mathbf{k}}$ are expected to be essential for the field pointing out of plane.
In fact such triplets potentially play an important role in the field induced transition to the odd parity singlet state in the locally noncentrosymmetric CeRh$_2$As$_2$ superconductor \cite{Mockli2021}.

The prescription for obtaining the triplet contribution [Eqs.~\eqref{eq:deltaL1_triplet} and \eqref{eq:deltaL2}] is in a similar vein to the singlet channel pairing in Appendix~\ref{app:pairing_s}. The triplet pairing Hamiltonian, Eq.\eqref{eq:Hpt}, is transformed to the basis of Eq.~\eqref{eq:psiB}. For the appropriate \( \mathbf{d}_\mathbf{k} \) directions for the Rashba SO interaction, Eq.~\eqref{eq:d-vector}, the Cooperon vertex
\begin{align}\label{eq:triplet_vertex}
	& \left[ i \check{\sigma}_x \left( \vb{k}, \vb{q}, \vb{B} \right) \right]_{ \lambda \lambda' }  
		=
	 \frac{1}{2} \left[
		\lambda
		\frac{ \left( k_y + \frac{  q_y }{2} + \frac{ B_x }{ \alpha } \right) - i \left( k_x + \frac{ q_x }{ 2 } - \frac{ B_y }{ \alpha } \right) }
		{ \sqrt{ \left( k_x + \frac{ q_x }{ 2 } - \frac{ B_y }{ \alpha } \right)^2 + \left( k_y + \frac{  q_y }{2} + \frac{ B_x }{ \alpha } \right)^2 } } 
		+
		 \lambda'
		\frac{ \left( - k_y + \frac{  q_y }{2} + \frac{ B_x }{ \alpha } \right) - i \left( - k_x + \frac{ q_x }{ 2 } - \frac{ B_y }{ \alpha } \right) }
		{ \sqrt{ \left(  k_x - \frac{ q_x }{ 2 } + \frac{ B_y }{ \alpha } \right)^2 + \left(  k_y - \frac{  q_y }{2} - \frac{ B_x }{ \alpha } \right)^2 } } 
		\right]
\end{align}
is almost the same as the singlet vertex, Eq.~\eqref{SC_Pauli_chiral}, except for the + sign prefactor for the \( \lambda \) term. There is a similar relation between the triplet Coopron vertex, \( \left[ i \check{\sigma}_x \left( \vb{k}, \vb{q}, \vb{B} \right) \right]_{ \lambda \lambda' }^\dagger \) , and the singlet \( \left[ i \check{\sigma}_y \left( \vb{k}, \vb{q}, \vb{B} \right) \right]_{ \lambda \lambda' }^\dagger \), Eq.~\eqref{SC_Pauli_chiral_hc}.

For the triplet corrections we expand the vertices to fourth order in \( \vb{Q}^\lambda \) [Eq.~\eqref{eq:Q}]:
\begin{subequations}
    \begin{align}
    	\left[i \check{\sigma}_x \left( \varphi, \vb{q}, \vb{B} \right) \right]
    		= & i e^{ i \varphi } \sigma_x
    		+
    		\frac{ e^{ i \varphi } }{ k_F } \left[ Q_y \cos(\varphi) - Q_x \sin(\varphi) \right] \sigma_z
    	\notag \\
			& + 
			\left[
                3 e^{ 3 i \varphi } \left( Q_x - i Q_y \right)^2 - e^{ - i \varphi} \left( Q_x - i Q_y \right)^2 - 2 e^{ i \varphi } \left( Q_x^2 + Q_y^2 \right) 
            \right] \frac{i \sigma_x }{8 k_F^2} 
    		+ T_3 \left( \vb{Q}^3 \right) \sigma_z + T_4 \left( \vb{Q}^4 \right) \sigma_x
    \label{eq:CooperT1}
    \\
    	\left[i \check{\sigma}_x \left( \varphi, \vb{q}, \vb{B} \right) \right]^\dagger
    		& = - i e^{ - i \varphi } \sigma_x
    		+
    		 \frac{ e^{ - i \varphi } }{k_F} \left[ Q_y \cos(\varphi) - Q_x \sin(\varphi)  \right] \sigma_z
    	\notag \\
    	+ &  
        \left[
            e^{ i \varphi } \left( Q_x - i Q_y \right)^2 - 3 e^{ - 3 i \varphi} \left( Q_x - i Q_y \right)^2 + 2 e^{ - i \varphi } \left( Q_x^2 + Q_y^2 \right) 
        \right]
    	\frac{ i\sigma_x }{8 k_F^2} 
    	+ T_3^d \left( \vb{Q}^3 \right) \sigma_z + T_4^d \left( \vb{Q}^4 \right) \sigma_x
    \label{eq:CooperT1dag}
    \end{align}
\end{subequations}
with \( T_3 \left( \vb{Q}^3 \right), T_3^d \left( \vb{Q}^3 \right), T_4 \left( \vb{Q}^4 \right), T_4^d \left( \vb{Q}^4 \right) \) encompassing the third- and fourth-order expansions that do not appear in the final result. The exact value of these terms is cumbersome and does not aid in understanding the calculation and is thus omitted.

Substituting Eqs.~\eqref{eq:CooperT1} and \eqref{eq:CooperT1dag} into the triplet version of Eq.~\eqref{eq:Bubble_w_Cooperon} gives
\begin{align}
    \Pi_t = \Pi_t^{ \left( 2 \right) } + \Pi_t^{ \left( 4 \right) } \,,
\end{align}
with the superscript indicating the order in \( \vb{Q}^\lambda \).

The correction to \( \mathcal{L}_1 \) is normalized with respect to the singlet pairing zero order [Eq.~\eqref{eq:Cooper_log}]
\begin{align}\label{eq:delta_L1t_calc}
	\frac{\Pi_t^{\left( 2 \right)}}{\Pi_s^{\left( 0 \right)} \left( T_{c0} \right)  } 
		= - 4 \frac{ \left( \vb{q} \cross \vb{B} \right) \cdot \hat{z} }{ k_F \Delta_{\mathrm{SO}} } \frac{g_t}{g} \,.
\end{align}
The triplet correction to the Lifshitz invariant~\eqref{eq:delta_L1t_calc} gives the same correction as singlet correction, Eq.~\eqref{eq:delta_L1s_calc}, up to numerical factor and the ratio of relative pairing strengths of the triplet and the singlet channels, \( g_t / g \), with the singlet pairing strength related to the Cooper logarithm, Eq.~\eqref{eq:Cooper_log}. This gives the correction \( \delta \mathcal{L}_1^t \) in Eq.~\eqref{eq:deltaL1_triplet}.

The triplet corrections to \( \mathcal{L}_2 \),
\begin{align}\label{eq:delta_L2t_calc}
	\frac{\Pi_t^{\left( 4 \right)}}{\Pi_s^{\left( 0 \right)} \left( T_{c0} \right) }
		= 
		- \frac{1}{2}
		\left[ 
			\frac{ \left( \vb{q} \cdot \vb{B} \right)^2 - \left( \vb{q} \cross \vb{B} \right)^2 }{ k_F^2 \Delta_{\mathrm{SO}}^2 \Lambda } + \frac{q^2 B^2}{2 k_F^2 \Delta_{\mathrm{SO}}^2 \Lambda } 
	    \right]
	    \frac{g_t}{g}
	\,,
\end{align}
are the same as these for singlet, Eq.~\eqref{eq:delta_L2s_calc}, up to the ratio of pairing strengths, \( g_t/g \ll 1 \). 
Equation~\eqref{eq:delta_L2t_calc} immediately gives the second term of Eq.~\eqref{eq:deltaL2}. Note that we neglect small terms such as \(  B^2/k_F^2 \) which arise in this calculation.

\end{widetext}

%\bibliography{Planar_Hall_effect_with_SC_fluctuations_bib}


\end{document}